\documentclass[aps,pra,10pt, twocolumn,showpacs,preprintnumbers,amsmath,amssymb, longbibliography]{revtex4-1}
\usepackage{graphicx}
\usepackage{siunitx}
\usepackage{braket}
\usepackage[nolist]{acronym}
\usepackage{multirow}
\usepackage{dcolumn}
\usepackage[dvipsnames]{xcolor}

\newcommand{\ra}[1]{\renewcommand{\arraystretch}{#1}}	
\newcommand{\head}[1]{\multicolumn{1}{c}{#1}}

\newcommand{\rcomp}{\mathbb{C}\backslash\mathcal{R}}
\usepackage{hyperref}

\newcommand{\enot}[2]{{#1}\!\times\!10^{#2}}
\newcommand{{\systemB}}{QDS-$b$-QSS-$b$-CV-QPSK}
\newcommand{{\systemF}}{QDS-$f$-QKD-$f$-CV-QPSK}

\usepackage{color}

\begin{document}

\begin{acronym}
\acro{QKD}[QKD]{quantum key distribution}
\end{acronym}

\author{Stefan Richter$^{1,2*}$, Matthew Thornton$^{3*}$, Imran Khan$^{1,2}$, Hamish Scott$^{3}$, Kevin Jaksch$^{1,2}$, Ulrich Vogl$^{1,2}$, Birgit Stiller$^{1,2}$, Gerd Leuchs$^{1,2}$, Christoph Marquardt$^{1,2}$, Natalia Korolkova$^{3}$}
\affiliation{$^1$Max Planck Institute for the Science of Light, Staudtstra{\ss}e 2, 91058 Erlangen, Germany}
\affiliation{$^2$Institute of Optics, Information and Photonics,
University of Erlangen-Nuremberg, Staudtstra{\ss}e 7/B2, Erlangen, Germany}
\affiliation{$^3$School of Physics and Astronomy, University of St Andrews,
North Haugh, St Andrews KY16 9SS, UK}
\email{stefan.richter@mpl.mpg.de, mt45@st-andrews.ac.uk}

\thanks{These authors contributed equally to this work.}


\title{Agile and versatile quantum communication: signatures and secrets}
\begin{abstract}
Agile cryptography allows for a resource-efficient swap of a cryptographic core in case the security of an underlying classical cryptographic algorithm becomes compromised. Conversely, versatile cryptography allows the user to switch the cryptographic task without requiring any knowledge of its inner workings. In this paper, we suggest how these related principles can be applied to the field of quantum cryptography by explicitly demonstrating two quantum cryptographic protocols, quantum digital signatures (QDS) and quantum secret sharing (QSS), on the same hardware sender and receiver platform. Crucially, the protocols differ only in their classical post-processing. The system is also suitable for quantum key distribution (QKD) and is highly compatible with deployed telecommunication infrastructures, since it uses standard quadrature phase shift keying (QPSK) encoding and heterodyne detection.  For the first time, QDS protocols are modified to allow for postselection at the receiver, enhancing protocol performance. The cryptographic primitives QDS and QSS are inherently multipartite and we prove that they are secure not only when a player internal to the task is dishonest, but also when (external) eavesdropping on the quantum channel is allowed. In our first proof-of-principle demonstration of an agile and versatile quantum communication system, the quantum states were distributed at GHz rates. This allows for a one-bit message to be securely signed using our QDS protocols in less than $0.05$~ms over a $2$~km fiber link and in less than $0.2$~s over a $20$~km fiber link. To our knowledge, this also marks the first demonstration of a continuous-variable direct QSS protocol.
 \end{abstract}

\maketitle

\section{Introduction}

Throughout history, cryptography has been threatened by advances in mathematics, computational power and side-channel attacks, and may soon be threatened by quantum computers. The breaking of a cryptosystem, i.e. a suite of cryptographic algorithms and hardware needed to implement a particular security service, has usually triggered the development of new algorithms. These would subsequently be tested and hardened for years before they could finally be deployed into real world applications to secure our ever-growing digital infrastructure. The redeployment of cryptographic soft- and hard-ware is a costly endeavor.

In the past decade, crypto-agility has emerged as a prospective solution to this problem \cite{sullivan_2009, *nist_report_2016}. One of the core ideas of crypto-agility is to provide a middleware with a two-way interface between the software application layer and the crypto-core or algorithm of the cryptosystem, Fig.~\ref{fig:agility}~(a), so that whenever a new attack vector emerges, the deployed architecture may stay in place and only the vulnerable crypto-core is replaced. This saves valuable deployment time as well as costs to re-engineer the whole system. The technical challenge is to design the middleware flexible enough to support novel crypto-cores. 

We here suggest how crypto-agility---and the related concept of crypto-versatility, in which multiple cryptographic tasks are performed on the same system---can be translated to quantum communication. Just as the quantum computer hardware provides qubits and gates to run different quantum algorithms on it, we propose that quantum communication hardware may support a diverse range of quantum communication protocols. By providing an abstraction layer between quantum-enabled hardware and the post-processing stack necessary to realize a quantum communication protocol, quantum versatility can be achieved. For our system this also implies quantum agility.

In this paper we explore agile and versatile quantum communication, and present an experimental demonstration of the first ``seed'' system featuring quantum crypto-agility and versatility. Specifically, we investigate continuous-variable quantum digital signatures (CV QDS), quantum secret sharing (CV QSS) and quantum key distribution (CV QKD) on a common platform. The secure protocols comprising the agile and versatile system use standard telecom sender and receiver techniques, thereby making the system both immediately compatible with deployed infrastructures, be it fiber networks or free-space links, and capable of high sending rates. Quantum coherent states, randomly chosen from an alphabet of four possible phases, are sent through a fiber-optic link, and highly efficient heterodyne detection is used at the receiver.

This first proof-of-principle agile and versatile quantum communication system is thus capable to perform three different quantum cryptographic protocols, QDS, QSS, and QKD using the same sending and receiving hardware for all protocols. The employed physical system and the advances made in the security proofs of the protocols allow for an implementation compatible with telecom networks. Along with the agility and versatility aspects, this work marks the first demonstration of our CV QSS scheme and the first demonstration of a CV QDS system with GHz sending rates and record speed to sign a one-bit message. Our demonstration thus provides a step towards a full quantum crypto-agility and versatility, in which several different quantum cryptographic protocols may be implemented on the same hardware deployment with alterations only at the level of classical postprocessing.

Our paper is outlined as follows. In Sec.~\ref{sec:2}, we propose and discuss two alternative approaches toward quantum crypto- agility and versatility, show that existing trends in the QKD and QDS literature may be interpreted in each context, and provide practical indications for when a quantum system may be deemed either agile or versatile. In Sec.~\ref{sec:3} we discuss three cryptographic tasks, QDS, QSS and QKD, and introduce several secure protocols which rely on the same physical setups. These protocols are implemented in Sec.~\ref{sec:4} and the resulting key rates and figures of merit are displayed in Sec.~\ref{sec:5}. We believe this demonstrates a crucial proof-of-principle step towards full quantum crypto-agility and crypto-versatility. Finally, we discuss our achievements through the lens of agility and  versatility in Sec.~\ref{sec:6}. 


\section{Quantum crypto-agility and versatility}\label{sec:2}

\begin{figure}[htp]
\centering
\includegraphics[width=\linewidth]{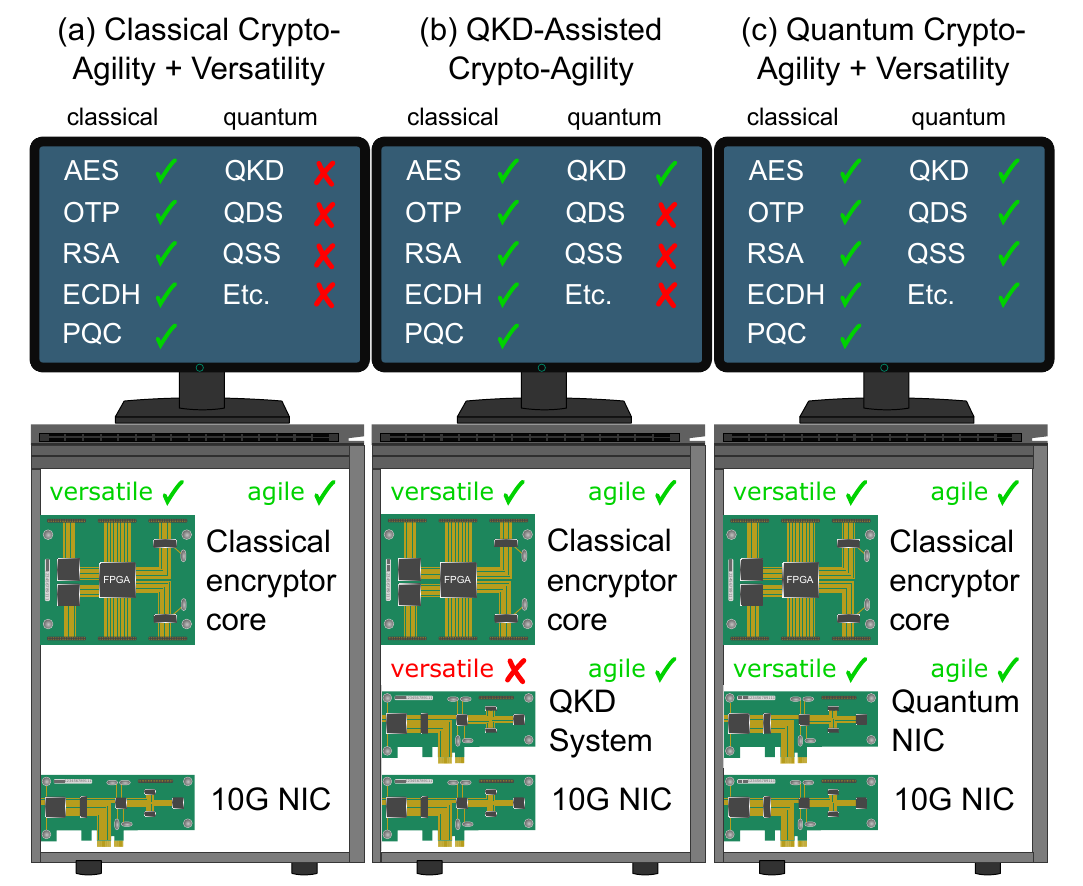}
\caption{
Comparison between classical crypto-agility and the proposed agile and versatile quantum cryptography architecture. 
(a) Classical Crypto- Agility and Versatility: Different classical cryptographic algorithms, such as Rivest-Shamir-Adleman\ (RSA), Elliptic-Curve Diffie-Hellman (ECDHE), Advanced Encryption Standard (AES), One-time Pad (OTP) or Post-Quantum Cryptography (PQC) can be flexibly combined on the same hardware platform. A network interface card (NIC) is used to send and receive secure communication.
(b) QKD-Assisted Crypto-Agility:  Classical cryptographic algorithms make use of a pool of secret keys generated by a QKD crypto core. The quantum functionality is tied to the hardware implementation and cannot easily be upgraded, e.g. to perform quantum digital signatures (QDS) or quantum secret sharing (QSS).
(c) Quantum Crypto-Agility: Classical cryptographic algorithms  and different quantum protocols can be swapped out and combined as necessary, requiring no changes to the underlying hardware architecture.
}
\label{fig:agility}
\end{figure}
Classical crypto- agility and versatility are described pictorially in Fig.~\ref{fig:agility}(a), in which a potentially vulnerable crypto-core may be readily replaced without affecting the rest of the deployed system, and in which several tasks can be accomplished via the same encryptor-core. The encrypted communication is then sent on the hardware level via a network interface card (NIC). The exact algorithm chosen to accomplish the task can be swapped and patched without knowledge of the end-user. We suggest here two different approaches to consider a {\it quantum} crypto system as agile or versatile. 

One can think of a first type of agility as {\it classical crypto-agility assisted by QKD}. Here a QKD system acts as a black box that delivers fresh shared keys to classical cryptography applications, see Fig.~\ref{fig:agility}(b). The advantage is that the middleware does not have to care about key generation or the QKD protocol itself. The downside is that although the generated key can be used for many different tasks, the QKD system itself may not be repurposed to run any other quantum protocol on it, limiting its versatility and potentially imposing an additional resource overhead.

\begin{figure}[htp]
\centering
\includegraphics[width=\linewidth]{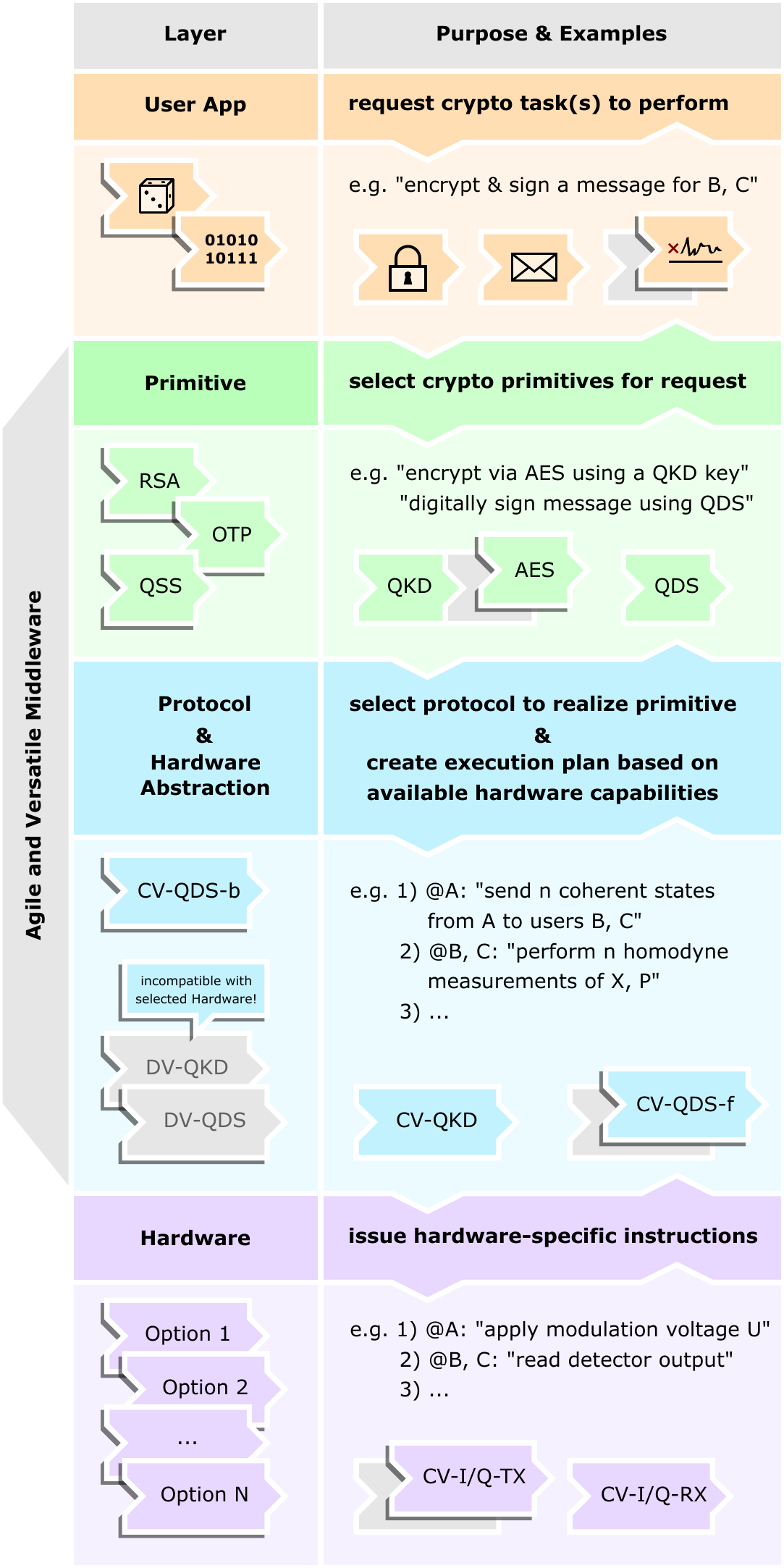}
\caption{A layer-based description of an agile and versatile quantum cryptosystem, showing how its modular and decoupled components can be swapped out and recombined, similarly to puzzle pieces. Quantum crypto-agility and versatility can both be realized by introducing a middleware (a collection of interface layers) between the user application (yellow) and quantum hardware (purple) layers. This requires that the hardware drivers expose a set of standardized functions to the layers above it. The middleware can then select suitable quantum crypto primitives and hardware-compatible protocols to fulfill a given user request for a given cryptographic task. In this manner, the middleware layers generalize and extend the functions of a key management system (KMS). For some of the acronyms occurring in this figure, please refer to the caption of Fig.~\ref{fig:agility}. CV-I/Q-TX and CV-I/Q-RX denote sender (Tx) and receiver (Rx) hardware modules capable of performing continuous-variable (CV) I/Q modulation and detection.}
\label{fig:layers}
\end{figure}

The second approach, {\it quantum crypto-agility + versatility}, is depicted in Fig.~\ref{fig:agility}(c). Compared to the first approach, the QKD system is replaced by a quantum network interface card (QNIC). The QNIC is able to perform multiple quantum communication protocols (e.g. QDS, QSS or QKD) on the same hardware platform. It communicates its hardware capabilities through an interface to the protocol layer, where the matching protocol for the user task at hand is chosen.  Such a layer stack is illustrated in Fig. \ref{fig:layers} and demonstrated later in this Paper. Note that here the choice of a particular quantum cryptographic application is reduced merely to a software and/or firmware update. A quantum crypto-system structured like this carries direct analogy with the classical agile crypto-system of Fig.~\ref{fig:agility}: hardware and agile interface stay the same and only the crypto-core, classical, as in Fig.~\ref{fig:agility}(a), or quantum, as in Fig.~\ref{fig:agility}(c), changes. This second approach to agility can be thought of as {\it a choice of quantum ``app''} and therefore also allows versatile usage of the quantum hardware. 

This approach carries an advantage of economic use of resources. QKD requires resource-intensive post-processing to generate a secure key, and real channel parameters (e.~g.,~noise, losses) may be too restrictive to allow for efficient secret key distillation. Some tasks, however, can be performed directly without first generating a shared secure key via QKD. A good example are QDS protocols, in which a secure signature is created straight from a raw quantum state exchange, consuming fewer resources  than an equivalent QKD protocol \cite{amiri_secure_2016}. Thus, a versatile system capable of performing both QDS and QKD will in general allow for a more efficient use of quantum resources when full QKD is either not possible or not necessary.

To make explicit the ideas discussed above, we propose the terms quantum crypto-agility and quantum crypto-versatility to mean the following (see. Fig.~\ref{fig:layers}): 
\\
\textbf{Quantum crypto-agility:} As a minimal requirement, a quantum crypto system can be considered agile if it exposes its cryptographic capabilities through a stable and opaque interface to the end user, allowing a compromised implementation to be modified without requiring changes to user software. A given crypto system may be agile up to different degrees, depending on which implementation components can be modified cost-effectively after deployment. It is typically easy to update a system's software, harder to update or replace firmware and quite difficult and costly to replace hardware. The transfer of the crypto-agility idea to the field of quantum cryptography slightly changes its meaning since the security of the quantum crypto-core can be proven information-theoretically. However, considering practical implementation security and the emergence of novel quantum cryptographic protocols and performance improvements to existing ones, an agile strategy seems prudent. 
\\
\textbf{Quantum crypto-versatility:} We consider a quantum system to be versatile if it implements different cryptographic tasks (e.g. QKD, QDS, QSS) on the same hardware platform and makes these available to the user through a common interface. In a versatile quantum crypto system, the selection of a quantum protocol to fulfill the requested cryptographic task happens on the middleware level, Fig.~\ref{fig:layers}. Since the inner workings of the specific task is only of concern for the manufacturer but not the user, this also implies agility. The converse may not hold true.

Quantum crypto-agility and versatility may also be relevant topics for ongoing standardization efforts, such as the ETSI QKD ISG 004 and 014 standards \cite{etsi_isg_004, etsi_isg_014} that define the interface between applications and key providers such as a key management system or QKD systems. Based on these two standards, quantum crypto-agility and versatility could eventually be added as another standard in a lower abstraction layer to form a full stack in the future. 

The idea of a layered architecture for QKD-secured communication systems is a natural one, and has been investigated before, e.g. in Refs. \cite{tysowski_engineering_2017, dahlberg_link_2019, kozlowski_2019, kozlowski2020designing}. For example, large quantum networks of quantum senders and receivers, classical communication lines and trusted/untrusted central nodes have been considered in Ref.~\cite{tysowski_engineering_2017}, where the study of different network topologies, and of the relationships between existing classical and future quantum networks, is important. Further, quantum analogues to the TCP/IP stack are discussed in Refs.~\cite{kozlowski_2019, kozlowski2020designing}, and quantum repeater links are also considered, allowing for additional teleportation-based protocols over the network. In contrast to these outlooks, our focus is on the layered stack required for individual nodes in the network, and is complementary to these full-network approaches. Notably, to our knowledge, neither CV nor DV based systems capable of selectively performing several different quantum primitives on the same hardware have been demonstrated so far.

In the remainder of this paper we will thus demonstrate the versatility of two quantum systems by showing that the choice of quantum cryptographic task can be made entirely at the software level. Because of the employed middleware, which provides an innate separation between Hardware and User layers, this also implies that our system is agile.

\section{Beyond QKD: Signatures and secrets}\label{sec:3}

In addition to the usual bipartite QKD, and in order to make our notion of quantum crypto-agility and versatility concrete, we consider the following multipartite tasks:

{\bf{QDS - quantum digital signatures}}: allows for the secure authentication of a classical message. It has been shown that because of its small overhead, QDS may run over channels for which QKD is insecure \cite{amiri_secure_2016}.

{\bf{QSS - quantum secret sharing}}: allows for the secure distribution of a classical secret among a conspiracy of potentially dishonest recipients.

\begin{figure}[htp]
\centering
\includegraphics[width=\linewidth]{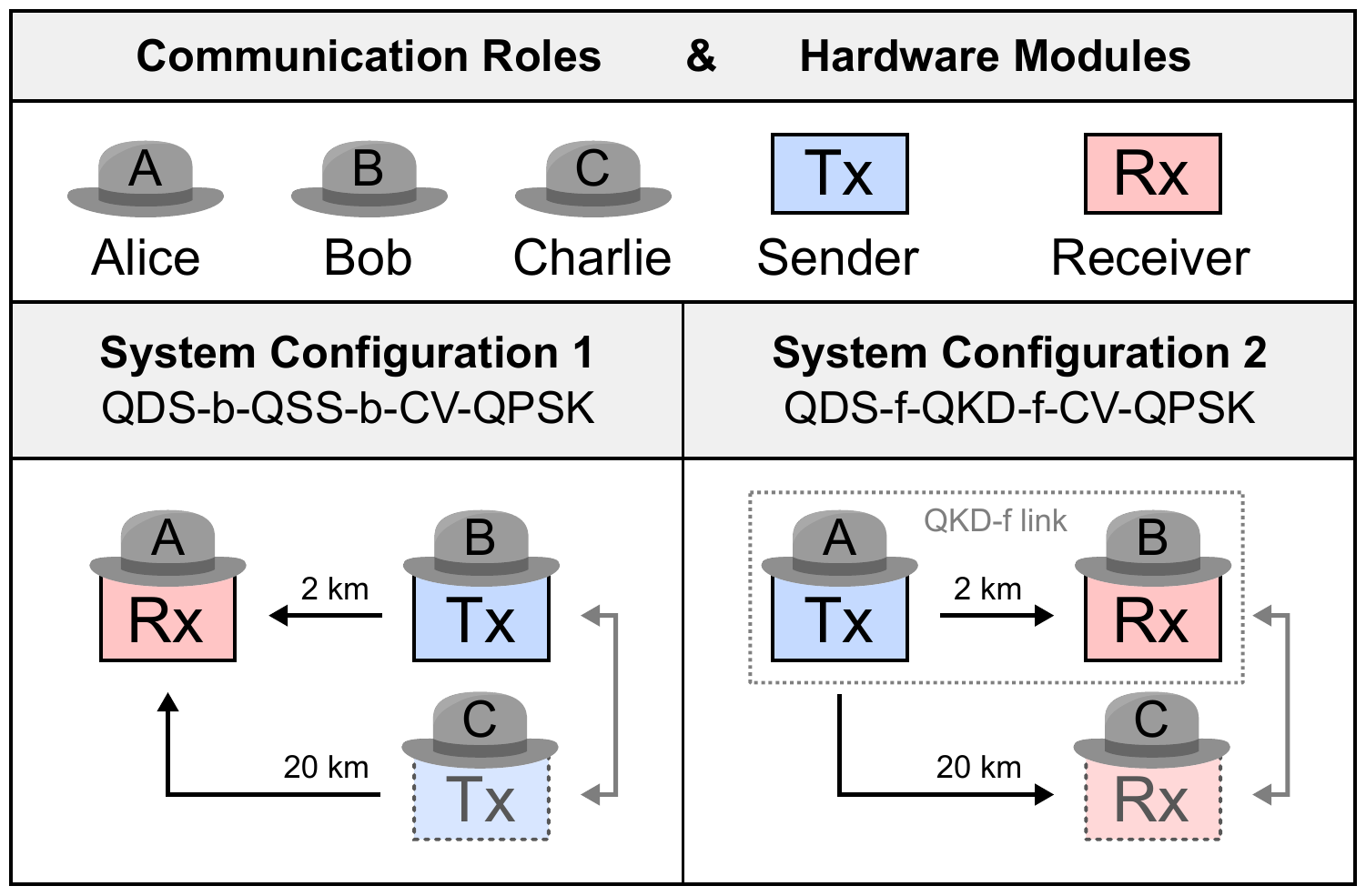}
\caption{
The sender (Tx) and and receiver (Rx) modules may be reconfigured to make Alice either the sender (``$f$-configuration'') or receiver (``$b$-configuration'') of quantum states. Each setup may be immediately considered \textit{versatile}, since once either the $f$- or $b$-configuration is chosen, multiple different cryptographic tasks may be performed. \textit{Agility} is implied by separation of the User and Hardware layers through a stable and opaque interface (see Fig.~\ref{fig:layers}). Shading and a dashed outline are used to indicate that a single device may be used to emulate two different endpoints as described in Sec. \ref{sec:4}.
}
\label{fig:txrx_roles} 
\end{figure}

In the spirit of quantum crypto-agility and versatility introduced earlier, Figs.~\ref{fig:agility}~(c),~\ref{fig:layers}, we explicitly propose two communication systems, i.e. configurations of the same underlying hardware, which can each fulfill multiple quantum cryptographic tasks. The two systems may thus both be considered as agile and versatile, and we denote them {\systemB} and {\systemF}, see Fig.~\ref{fig:txrx_roles}. The labels indicate which cryptographic tasks (QDS, QSS or QKD) they support; the underlying quantum states that they use (a continuous-variable CV-QPSK alphabet); and in which direction ($f$ - ``forward'' or $b$ -``backward'') the quantum states are exchanged. Labeling agile and versatile quantum cryptosystems by the hardware components they are based on and the protocols they support might prove useful in later efforts to standardize interfaces and provide some comparability between different implementations.

The agile and versatile approaches can in principle be applied to both discrete- and continuous- variable systems, with the agile middleware, Fig.~\ref{fig:layers}, ensuring that the end-user does not need to care about whether (quasi)-single photons or phase-encoded coherent states are used. For the remainder of the paper we focus on the CV platform, noting that the use of the QPSK alphabet and heterodyne detection renders our system highly compatible with standard telecom infrastructure, potentially paving a way to integrating agile and versatile quantum crypto systems into deployed communication links which run with up to $100$~GHz sending rate \cite{khan_continuous-variable_2015,khan_towards_2015}. With this in mind, the protocols presented here sit within the field of continuous-variable (CV) quantum cryptography, which aims towards fast sending rates over metropolitan distances. The four individual protocols each provide asymptotic security against a dishonest player performing a collective beamsplitter or entangling-cloner attack. Descriptions of each protocol and key details in their security proofs are sketched below, while the reader is referred to the appendices for technical details.

\subsection{First agile and versatile system {\systemB}}

The first agile and versatile system we consider relies on the $b$-configuration, with Bob and Charlie as the senders (Tx) of quantum states, while Alice performs heterodyne detection (Rx), Fig.~\ref{fig:txrx_roles}. This {\systemB} system is capable of performing both QDS and QSS tasks via the protocols QDS-$b$ and QSS-$b$, which are described below. Our experiment, detailed in Section~\ref{sec:4}, also marks the first demonstration of CV QDS over insecure quantum channels.

\subsubsection{The QDS-$b$ protocol}

The very first QDS scheme was proposed in Ref.~\cite{gottesman_quantum_2001} and required a quantum memory. In the last two decades, discrete-variable (DV) QDS protocols have first lifted this requirement \cite{andersson_experimentally_2006, clarke_experimental_2012, dunjko_quantum_2014, collins_realization_2014} and then also have lifted the need for a trusted quantum channel \cite{amiri_secure_2016, yin_practical_2016}, and have brought their hardware requirements closer to those of QKD \cite{wallden_quantum_2015}. Recently, DV QDS implementations based on deployed networks have been demonstrated successfully over metropolitan distances \cite{collins_experimental_2016, yin_experimental_2017, yin_experimental_2017-1, an_practical_2019, Ding2020}. Indeed, in several QDS papers, a nascent form of quantum versatility is mentioned, either explicity \cite{collins_experimental_2016, yin_experimental_2017-1, roberts_experimental_2017} or implicitly \cite{yin_experimental_2017, an_practical_2019}, but so far the comparison has always been that the distribution of quantum states for QDS is analogous--or in some cases identical--to that required for QKD. For example, Ref. \cite{collins_experimental_2016} differs from differential-phase-shift QKD only in post-processing. Similarly, one protocol in Ref.~\cite{wallden_quantum_2015} is designed specifically to share sender and receiver with QKD, while another requires first full QKD and then classical communication to sign a message. Despite these recognitions, to our knowledge the full utility of applying the idea of quantum crypto-agility and versatility to a deployed quantum network has not yet been explored, nor have additional cryptographic protocols been studied in this framework.

Unlike those preceding QDS protocols, in which Alice was the sender of quantum states, in QDS-$b$ Bob and Charlie are the senders of quantum states, while Alice is the recipient, Fig.~\ref{fig:txrx_roles}~(b). This allows QDS-$b$ to be performed on our first agile and versatile system {\systemB}.

The protocol QDS-$b$ runs as follows:

\noindent 1) For each future message $m \in \left\{0, 1\right\}$, which Alice wishes to securely sign, Bob and Charlie both send Alice a sequence, length $L$, of coherent states chosen randomly from the QPSK alphabet. Bob and Charlie each keep a record of which states they have sent.

\noindent 2) Alice performs heterodyne detection on each received state, and forms eliminated signatures $A_{B, C}^m$ by writing down which two states from QPSK are \textit{least compatible} with her measurement outcome, that is, which states have the smallest conditional probability of being sent.

\noindent 3) Bob and Charlie swap a random half of their signature elements in order to guard against dishonest Alice 
\footnote{The exchange of eliminated signature elements in a QDS protocol must be kept hidden from Alice. This swapping is most straightforwardly done over an encrypted channel which implicitly requires QKD. To allow for a fair comparison to existing QDS schemes and implementations, we omit an accounting for this QKD link, instead using in further QDS analyses the figures of merit commonly used by the QDS community \cite{donaldson_experimental_2016, collins_experimental_2016, amiri_secure_2016, croal_free-space_2016, collins_progress_2018}}. Bob (Charlie) now possesses signatures  $X_B^m$, $\left(X_C^m\right)$, which consists of two halves of length $L/2$, one of which was generated by Bob (Charlie), and one of which was received during the swapping.

\noindent 4) Later, Alice sends her message $m$ and the corresponding $A_{B, C}^m$, first to Bob. Bob compares his record of which states were sent, and counts the number of mismatches. A mismatch occurs if Alice claims to have eliminated a state which Bob indeed did send. Provided that there are sufficiently few mismatches, he accepts $m$ as genuine and forwards to Charlie, who likewise accepts or rejects by counting the number of mismatches.

A QDS scheme must be secure against both \textit{forging} attacks, in which a dishonest Bob will attempt to convince Charlie that a message is genuine; and \textit{repudiation} attacks, in which a dishonest Alice will attempt to force Bob and Charlie to disagree about the message's validity. Furthermore, noting that a QDS protocol which declares all possible signatures as fake may be considered trivially secure, we require that the protocol should succeed if all parties are honest, that is, it should be \textit{robust}. 

The full security proof of QDS-$b$ may be found in Appendix~\ref{appendix:QDS2forge}. Here we simply note that security against forgery is guaranteed by picking a highly non-orthogonal alphabet of coherent states, i.e. the amplitude of the QPSK alphabet should be sufficiently small. 

The main security result for QDS-$b$ is the following expression for the binary entropy $h$
\begin{equation}\label{eqn:qdsbsecurity}
h\left(p_e\right) \ge 1 - \chi
\end{equation}
of a forging Bob's probability $p_e$ to induce a mismatch with Charlie. The $\chi$ denotes Bob's Holevo information about Charlie's distributed state. Then the final signature length $L$ required to sign a $1$~bit message with $\varepsilon_{fail}$ probability of failure is implicitly given by \cite{thornton_continuous-variable_2019, croal_free-space_2016}
\begin{equation}\label{eqn:QDSbL}
\varepsilon_{fail} \le 2 \exp \left[ - \frac{\left( p_e - p_{err} \right)^2}{16} L \right]
\end{equation}
provided that security parameter $p_e - p_{err} > 0$, where $p_{err}$ is an honest player's mismatch probability, which can be estimated during the protocol. In other words, QDS-$b$ is secure against any attack provided that a dishonest player causes more mismatches than an honest player.

The protocol QDS-$b$ performs well over channels with low loss and low excess noise, but in order to reach feasible signature lengths over realistic channels we will employ the \textit{postselection} technique \cite{Silberhorn2002}. To our knowledge this is the first time this technique has been leveraged in the context of QDS. Alice will discard measurement outcomes for which she has a large probability of mismatch, thereby reducing $p_{err}$. Since a forger will attack the sender (Tx) of quantum states rather than Alice, the probability $p_e$ is unaffected by postselection. The security parameter $p_e - p_{err}$ may then be readily altered simply by choice of postselection region. The full postselection calculation is found in Appendix~\ref{appendix:PS}.

The experimental implementation of the protocol is presented in Sec.~\ref{sec:4}, and the signature length $L$ required to sign a $1$~bit message to $\varepsilon_{fail}$ chance of failure is given in Sec.~\ref{sec:5}. 

\subsubsection{The QSS-$b$ protocol}

A secret-sharing scheme allows for Alice to distribute a classical secret between recipients Bob and Charlie. Bob and Charlie should be able to perfectly reconstruct the secret when they behave honestly, while either Bob or Charlie working alone should gain no information.

Although some existing \emph{classical} secret-sharing schemes are already information-theoretically secure \cite{Schneier1996}, they encounter problems when distributing the shares of the secret across insecure channels, and may fall prey to an eavesdropper with a sufficiently powerful quantum computer. A potential solution is to employ a quantum secret sharing (QSS) protocol which uses quantum resources in order to share the classical secret \cite{Kogias2017, Grice2019}. For example, the scheme put forward in Ref.~\cite{Kogias2017} relies on large multipartite entangled states for distillation of keys between the dealer, Alice, and a degree of freedom shared between recipients. In another protocol \cite{Grice2019} security is reached via a ``round-robin'' distribution stage with each player interacting with the same transmitted quantum state. 

Crucially, unlike these approaches which require dedicated hardware setups or distribution of large entangled states, the QSS-$b$ protocol presented here accomplishes the secret-sharing task using only distribution of QPSK coherent states and heterodyne detection, and thus forms an integral part of our first agile system {\systemB}, Fig.~\ref{fig:txrx_roles}. We demonstrate in Sec.~\ref{sec:5} that QSS-$b$ attains a larger key rate than an equivalent information-theoretically secure classical secret sharing scheme using two continuous-variable QKD setups.

In the QSS-$b$ protocol, the dealer (Alice) is assumed honest, while either one of the Bob or Charlie may be dishonest. Additionally, a dishonest fourth player, Eve, may be present. For now we assume that a dishonest Bob/Charlie will send states only from the QPSK alphabet, though this could be relaxed in future work.

The protocol QSS-$b$ runs as follows:

\noindent $1$) Bob and Charlie send sequences of coherent states to Alice, which are independently and randomly chosen from the QPSK alphabet. Alice performs heterodyne measurement of phase and records her outcomes $A_B, A_C \in \mathbb{C}$. Bob and Charlie keep a record, $X_B, X_C$, of which states they have sent. 

\noindent $2$) Alice forms a variable $X_A \simeq F\left(A_B, A_C\right)$ which is some function $F$ of her measurement results. She then encodes the secret using the $X_A$, and makes the encoded secret publicly available.

\noindent $3$) Later, when Alice wishes to allow Bob and Charlie to reconstruct the secret, she leaks the function $F$ and enough information to perform a reconciliation procedure between her $X_A$ and the $X_A \simeq F\left(X_B, X_C\right)$ generated by Bob and Charlie. The reconciliation proceeds as in regular QKD.

\noindent $4$) Bob and Charlie, by working together to form and reconcile $F\left(X_B, X_C\right)$, gain a copy of Alice's key. Thus they are able to decrypt her message.

The protocol should prevent dishonest players from reconstructing the secret unless they collaborate with the honest player. Specifically, they are forced to collaborate by Alice's choice of $F$ which requires information from both players to reach the key. The function $F$ can be arbitrarily chosen and optimized over, though for concreteness we choose $F$ to be linear, i.e.
\begin{equation}
F\left(X_B, X_C\right) = g X_B + h X_C
\end{equation}
for $g, h \in \mathbb{R}$; Alice is free to choose a more general $F$ if it is optimal for her setup.

The security proof for QSS-$b$ is found in Appendix~\ref{appendix:QSS}. The main security result is a calculation of the key rate $\kappa$ generated between Alice and a Bob-Charlie collaboration. The key rate corresponds to the number of secure key bits which which may be encrypted per channel use, i.e. after both Bob and Charlie have sent a state. One channel use thus corresponds to distribution of \emph{two} coherent states.

In the the presence of dishonest Eve and honest Bob/Charlie the key rate $\kappa$ is given by the following Devetak-Winter bound \cite{Kogias2017, Devetak2005}
\begin{equation}\label{eqn:QSS_eavesdropper}
\kappa \ge I\left(X_B, X_C : X_A \right) - \chi \left( X_A : \mathbb{E}\right)
\end{equation}

\noindent relating the mutual information $I$ between Bob/Charlie's classical information $X_{B, C}$ and Alice's information $X_A$, and the Holevo information $\chi$ which Eve's quantum system $\mathbb{E}$ holds about $X_A$. 

More general bounds to guard against dishonest Bob/Charlie are given in the Appendix. The QSS-$b$ protocol is implemented in Sec.~\ref{sec:4}, and the key rate to allow for secure secret-sharing is given in Sec.~\ref{sec:5}.

\subsection{Second agile and versatile system {\systemF}}
In addition to our first agile and versatile system described above, which is capable of readily switching between QDS and QSS tasks, we will demonstrate that cryptographic protocols which already exist in the literature may be viewed through an agility/versatility lens. We therefore turn to consider a second agile and versatile system, denoted {\systemF}, which is capable of performing either QDS or QKD tasks in a ``forward''-configuration, Fig.~\ref{fig:txrx_roles}.

A QDS protocol in which Alice sends quantum coherent states was previously considered in Ref.~\cite{thornton_continuous-variable_2019}. There, it is Bob and Charlie who form eliminated signatures, and check for mismatches between their eliminated signatures and Alice's declaration of which states she sent. We here denote this protocol QDS-$f$.

To go beyond \cite{thornton_continuous-variable_2019} we apply the postselection technique to QDS-$f$, which decreases the number of quantum states $L$ required to sign a message, particularly in the presence of channel noise. Since (in contrast to QDS-$b$) it is now Bob and Charlie who heterodyne, rather than Alice, both terms $p_e$ and $p_{err}$ now change with the choice of postselection region. The effects of postselection on QDS-$f$, and key steps from the security proof, are detailed in Appendix~\ref{appendix:PS}.	

Finally, we round-off the second agile and versatile system by noting that the discrete-modulation QPSK QKD protocol, analysed e.g. in Ref.~\cite{Papanastasiou2018}, may be readily implemented using the same hardware setup as QDS-$f$ without requiring reconfiguration. This protocol, which we here denote QKD-$f$, may be performed between either Alice-Bob or Alice-Charlie. The full security proof is found in Ref.~\cite{Papanastasiou2018}, and we display the estimated maximum rate of secure key generation for our system under QKD-$f$ in Section~\ref{sec:5}.

\section{Experiment}\label{sec:4}

\begin{figure}[ht!p]
\centering
\includegraphics[width=\linewidth]{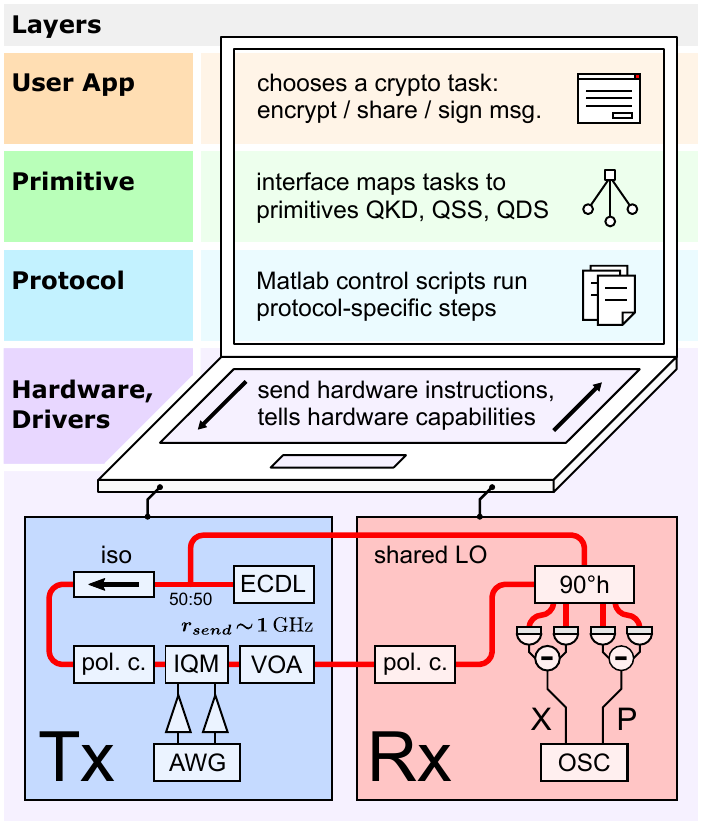}
\caption{Top: Schematic drawing indicating how the different abstraction layers of quantum crypto-agility and versatility (see Fig.~\ref{fig:layers}) are realized in our demo system as different pieces of software running on a laptop connected to our sender (Tx) and receiver (Rx)  hardware. Bottom: The Tx and Rx modules used to perform four distinct quantum protocols, assuming different communication roles A, B or C as shown in Fig.~\ref{fig:txrx_roles}. ECDL: external-cavity diode laser, iso: isolator, VOA: variable optical attenuator, IQM: I/Q modulator, pol.c.: polarization controller, AWG: arbitrary waveform generator, LO: local oscillator,  $90^{\circ}$h:  $90^{\circ}$hybrid, OSC: oscilloscope}.
\label{fig:txrx_exp}
\end{figure}

An optical sender (Tx) and receiver (Rx) module as shown in Fig.~\ref{fig:txrx_exp} are used to experimentally investigate the performance of the protocols QDS-$f$, QKD-$f$, QDS-$b$ and QSS-$b$.
Depending on the required configuration, Tx and Rx take on the roles indicated in Fig.~\ref{fig:txrx_roles}. Our protocols do not require the quantum state exchange between parties A-B and A-C to be simultaneous. For this demonstration, we thus emulate the deployment of identical Tx/Rx hardware at B and C by instead sequentially linking a single Rx and Tx using a \SI{2}{km} (\SI{-0.65}{dB}) or \SI{20}{km} (\SI{-4.75}{dB}) SMF-28 optical fiber channel. Each state exchange between A-B and A-C is  performed as an independent sub-experiment.

\begin{figure}[htp]
\centering
\includegraphics[width=\linewidth]{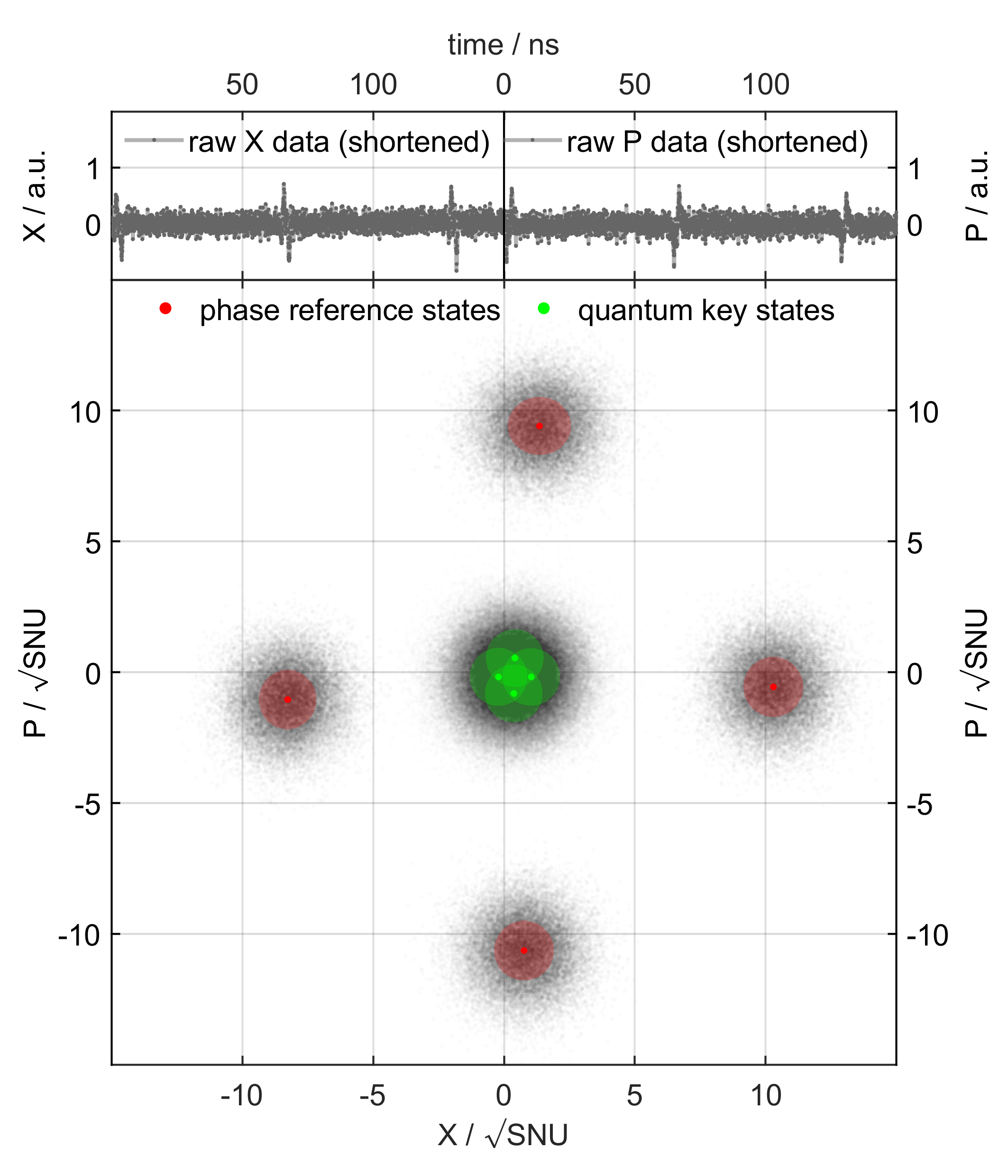}
\caption{
Top: Raw quadrature data traces produced by our quantum communication system running quadrature phase shift keying (QPSK) modulation.
Bottom: A resulting phase-space constellation diagram after digital signal processing (DSP) has been applied to the raw data. Shaded circles indicate the means and variances of the coherent states sent and received, including quantum key states (green) and auxiliary phase reference states (red).
\label{fig:expt_data} 
}
\end{figure}

\textbf{Sender module (Tx):}\ 

A PurePhotonics PPCL-300 external cavity diode laser (ECDL) with a linewidth of \SI{15}{kHz} tuned to a wavelength of \SI{1550}{nm} acts as an optical carrier. Using a Fujitsu DP-QPSK \SI{40}{Gbps} $LiNbO_3$-integrated I/Q modulator, driven at a rate of \SI{1}{GHz} by a Keysight M8195A arbitrary waveform generator (AWG), the sender randomly prepares pulses of coherent states chosen from the QPSK alphabet $\{|{\pm\alpha_0}\rangle, |{\pm i \alpha_0}\rangle\}$. These states are attenuated to a chosen output amplitude $\alpha < \alpha_0$ with a variable optical attenuator and sent to the receiver. 

\textbf{Receiver module (Rx):}\ The receiver module interferes the incoming signal with a local oscillator in an integrated Kylia COH24-X $90^\circ$ hybrid 
and performs heterodyne detection of the electric field quadratures $X$ and $P$ for each state using two Discovery DSC-R412 balanced optical receivers with an analog \SI{3}{dB}-bandwidth of \SI{20}{GHz}. For the purposes of this demonstration, the local oscillator is sourced from the carrier laser and transmitted to the receiver using an additional fiber.
The optical receiver outputs are digitized and processed on a Tektronix DPO77002SX digital sampling oscilloscope using a sampling rate of \SI{25}{GS/s}. Digital signal processing (DSP) is applied to the quadrature time traces, consisting of a high-pass filtering operation to eliminate low-frequency noise components and a phase recovery step using reference states. 

Experiments were performed for different modulation amplitudes $\alpha$, as indicated in Tab.~\ref{tab:lengths}. For each state exchange, a total of \num{1.92e6} states were sent in frames of 64, with 4 bright phase reference states at the start of each frame. Of those, \num{1.54e6} states or $\SI{80.2}{\%}$ remained after the DSP. A phase space diagram of the quantum state constellation and a section of the measured raw data can be seen in Fig. \ref{fig:expt_data}.

\section{Results}\label{sec:5}

\begin{figure}[htp]
\centering
\includegraphics[width=\linewidth]{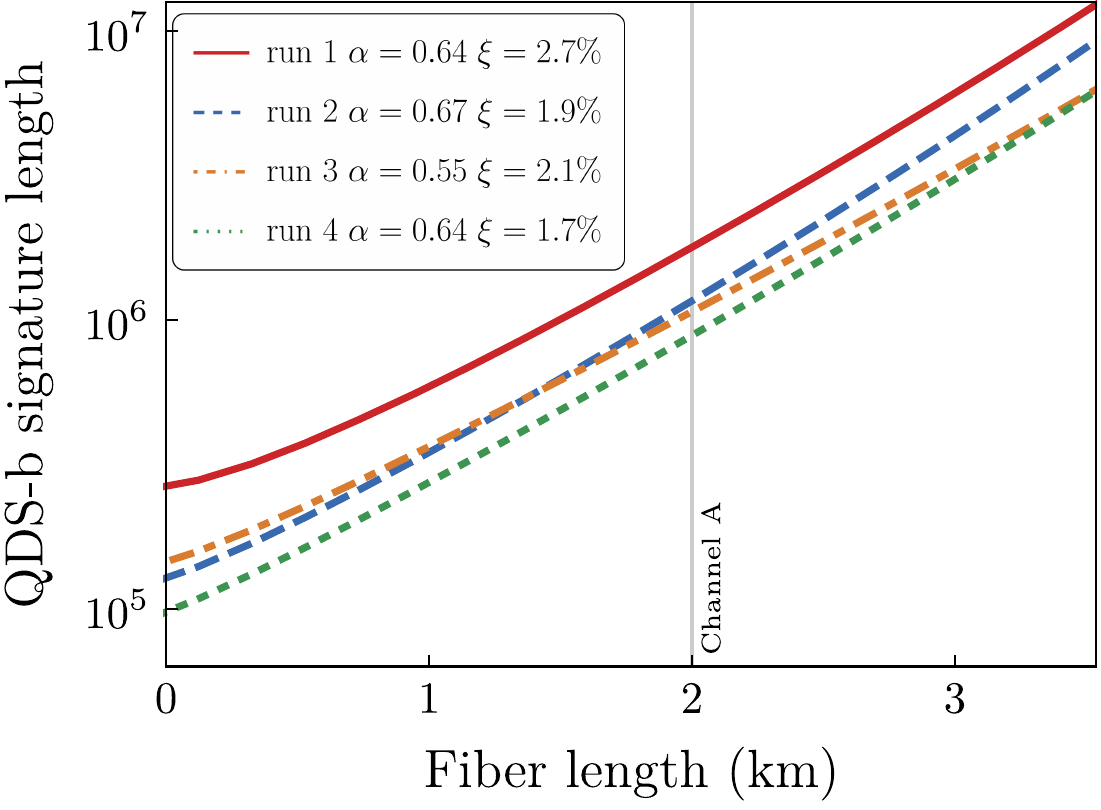}
\includegraphics[width=\linewidth]{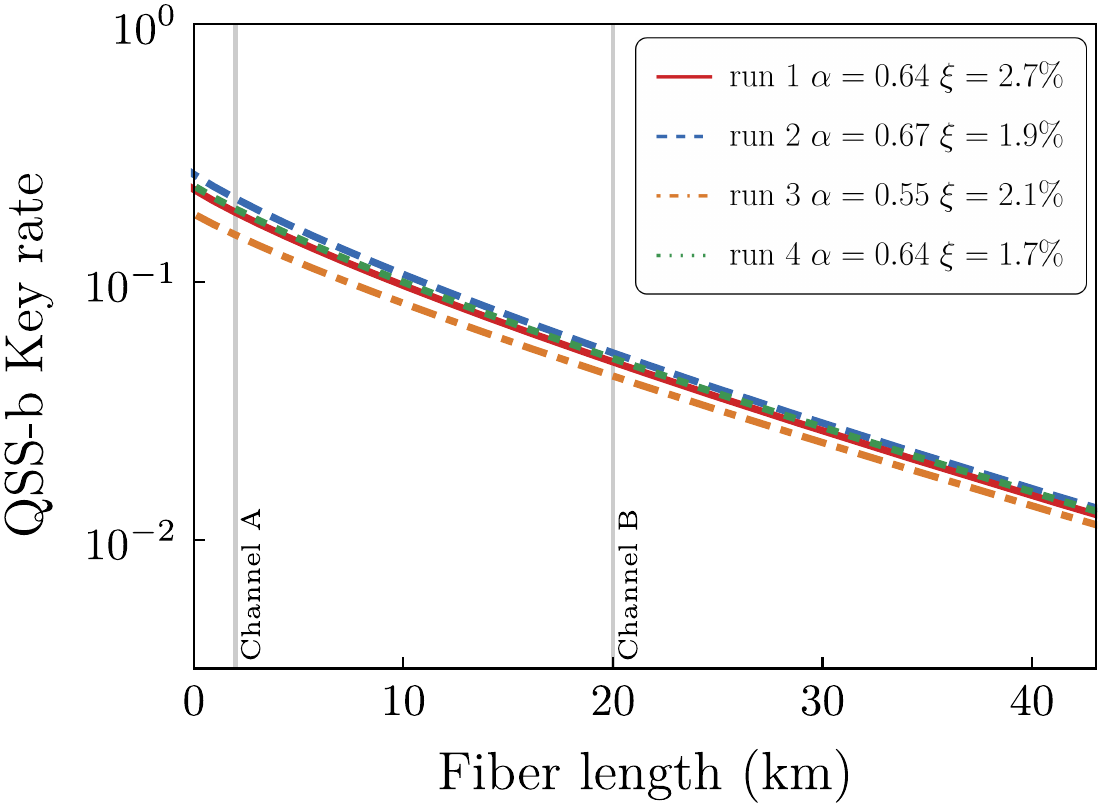}
\caption{\label{fig:qdsb_results} Performance of agile and versatile system {\systemB}. Top: QDS signature lengths under protocol QDS-$b$ with an entangling-cloner attack. The signature lengths at a distance of $2$~km remain modest both in the ideal (above) and experimental realizations (Tab.~\ref{tab:lengths}), and the system is robust to choice of $\alpha$. Bottom: Maximum calculated QSS key rates under protocol QSS-$b$ with a dishonest Eve performing a beamsplitter attack, and either Bob or Charlie dishonest. The key rate is robust to variations in $\alpha$, and remains large even for our $20$~km channel. Solid (red), dashed (blue), dot-dashed (orange) and dotted (green) lines correspond to the performance deduced by parameters from experimental runs $1$, $2$, $3$ and $4$, respectively. Vertical grid lines depict loss levels over experimental channels $A$ and $B$, corresponding to fiber lengths $2$~km ($0.65$~dB loss) and $20$~km ($4.75$~dB loss).}
\end{figure}

The agile and versatile system {\systemB} has been investigated over the $2$~km fiber link with average $\bar{\alpha}=0.64$ and an excess noise in the channel of \SI{2.7}{\%} in the laboratory conditions that represent the first targeted implementation of an agile and versatile quantum communication system. We obtain practical figures of merit for each of the protocols (5.7~ms to sign a 1-bit message for QDS-$b$ and $2\kappa = 0.3726$ key rate for QSS-$b$), listed in Tab.~\ref{tab:lengths}. The protocol QSS-$b$ has also been investigated in several $20$~km experimental runs for different $\alpha$ and different levels of excess noise with key rate up to $2\kappa = 0.1058$, completing the first demonstration of our practical CV QSS, Fig.~\ref{fig:qdsb_results}

The second agile and versatile system {\systemF} has been investigated over a $2$~km laboratory fiber link and in several runs over $20$~km for different amplitudes and different levels of excess noise in order to explore performance at larger distances, which are less favorable for CV communication systems. Alongside the agility and versatility aspects, this experiment has demonstrated the fastest to-date QDS system at intra-city distances, allowing to sign a $1$-bit message in less than $0.05$~ms over a $2$~km fiber link. It has also allowed for a secure performance of the agile system at $20$~km distance with feasible signature lengths, Fig.~\ref{fig:qdsf_results}, with signing times close to the recent best DV experiments, Fig.~\ref{fig:star}.

The maximum calculated secure key rates for QKD-$f$ for this agile and versatile system are displayed in Fig.~\ref{fig:qdsf_results} and Tab.~\ref{tab:lengths}. The maximum obtainable QKD key rate for the system is $\kappa = 0.1024$. 

We detail and benchmark the different aspects of the experimental performance of the two agile and versatile systems in what follows.

\begin{table*}
	\centering \ra{1.75}
	\begin{tabular*}{\textwidth}{@{\extracolsep{\stretch{1}}} cccc c rr c rr c r cc}
	\multicolumn{4}{c}{\textbf{Experiment}} &&
	\multicolumn{2}{c}{\textbf{QDS\,-\,$b$}} &&
	\multicolumn{2}{c}{\textbf{QDS\,-\,$f$}} &&
	\multicolumn{1}{c}{\textbf{QSS\,-\,$b$}} && 
	\multicolumn{1}{c}{\textbf{QKD\,-\,$f$}} \\
	\colrule
	\text{Run} & \head{$d\,[\si{km}]$} & \head{$\bar{\alpha}\,[\sqrt{\text{snu}}\,]$} & \head{$\xi\,[\si{\%}]$} &&
	\head{$L\,[\si{bits^{-1}}]$} & \head{$t\,[\si{ms}]$} &&
	\head{$L\,[\si{bits^{-1}}]$} & \head{$t\,[\si{ms}]$} &&
	\head{$2 \kappa$}  && 
	\head{$\kappa$}
	\\
	\colrule
	1 & \phantom{0}2 & 0.64 & 2.7 && $\enot{5.70}{6}$ & $5.7$ && $\enot{4.79}{4}$ & $0.048$ && 0.3726 && 0.3479\\
	2 &                   20 & 0.67 & 1.9 &&                          - &        - && $\enot{2.26}{9}$ &  $2260$ && 0.1058 && 0.1024\\
	3 &                   20 & 0.55 & 2.1 &&                          - &        - && $\enot{1.37}{8}$ &    $137$ && 0.0858 && 0.0840 \\
	4 &                   20 & 0.64 & 1.7 &&                          - &        - && $\enot{2.08}{8}$ &    $208$ && 0.1004 && 0.0976\\
	\end{tabular*}
	\caption{\label{tab:lengths} Figures of merit for the experimental runs. QDS signature lengths (L) and signing times (t) required to sign a $1$-bit message for security level of $\varepsilon = 0.01\%$. The QSS and QKD key rates correspond to the maximum estimated number of bits of secure key which may be generated per use of the quantum channel. In QSS-$b$, one channel use corresponds to distribution of \emph{two} quantum states, one from Bob and one from Charlie, and so we display $2 \kappa$ for fair comparison with QKD.}
\end{table*}

\subsection{Settings for the system runs}
We performed the experiment detailed above over two different channels which we denote channel $A$ and channel $B$, corresponding to $2$~km fiber length and $20$~km fiber length, respectively. During the experiment, measurement outcomes corresponding to parameters detailed in Tab.~\ref{tab:lengths} were obtained. Each element of the QPSK alphabet had slightly different sending amplitude in each experiment, and we display the average amplitude $\bar{\alpha}$ in the table. The excess noise $\xi$ above the shot-noise was calculated for each quadrature $x, p$ and $\xi = \text{max}\{\xi_x, \xi_p\}$ was taken as a worst-case scenario. We now process our measured data with reference to each of the four quantum protocols and thus demonstrate quantum crypto-versatility, thereby implying agility, for our systems.

\subsection{First agile and versatile system {\systemB}}

In the first agile and versatile system {\systemB}, the sender module Tx is understood to play the role of either Bob or Charlie, while Rx plays the role of Alice. 

\paragraph{QDS-$b$:} Signature lengths are calculated using data parameters from Tab.~\ref{tab:lengths} with the postselection region $\mathcal{R}\left(\Delta_r\right)$ optimized at each channel loss, see Appendix.~\ref{appendix:PS}. In the ideal case, the probability $p_{err}$ is calculated using Eq.~(\ref{eqn:perrPS}) under the model described in Appendix~\ref{appendix:PS}, which includes both channel excess noise $\xi$, ascribed to Eve, and a detector efficiency of $50\%$ which Eve cannot exploit. We allow Eve to perform the entangling cloner attack \cite{thornton_continuous-variable_2019} which is expected to be optimal in the limit $\alpha \rightarrow 0$, and close to optimal for the small $\alpha$'s used here, and probability $p_e$ may be estimated as in Appendix~\ref{appendix:QDS2forge} once the worst-case $\alpha$ and $\xi$ have been estimated from data. The ideal signature lengths for QDS-$b$ are displayed in Fig.~\ref{fig:qdsb_results}.

More realistic signature lengths may be calculated by taking into account in the estimate of $p_e$ the actual amplitudes and sending probabilities which Tx sent, rather than an average, and by measuring $p_{err}$ directly from the output of Rx. The $p_{err}$ calculated this way takes into account all sources of detector loss and trusted noise which will increase $p_{err}$, and thus the measured $L$ will be larger than those in Fig.~\ref{fig:qdsb_results}. 

Further, in comparisons between commercial implementations, the implicit Bob-Charlie QKD channel for the swapping stage of the QDS protocol should be included in a figure of merit to give a realistic accounting of the resource requirements. We stress however that the aim of our QDS analysis is primarily to give a comparison with recent QDS schemes (see Fig.~\ref{fig:star}). Therefore, our approach to the classical postprocessing focuses on the QDS distribution stages (following the QDS literature \cite{donaldson_experimental_2016, collins_experimental_2016, amiri_secure_2016, croal_free-space_2016, collins_progress_2018}) in order to allow for this comparison.

For experimental run $1$ over the $2$~km channel under entangling-cloner attack, signature length $L = 5.7\times10^6$ is required to sign a single bit, Tab.~\ref{tab:lengths}. However, even at $20$~km, QDS-$b$ could still be made secure by choosing a large postselection region with $\Delta_r \gg 1$, but for loss levels more than $\sim 2$~dB the signature length required becomes impractically large.

\paragraph{QSS-$b$:} For our secret-sharing protocol QSS-$b$, Fig.~\ref{fig:qdsb_results}, the Holevo information is calculated by estimating channel transmission $T$ and excess noise $\xi$ from the data and assuming the dishonest players perform a beamsplitter attack. In our reported results we have optimized over $g, h \in \mathbb{R}$ which parametrise the function $F\left(X_B, X_C\right) = g X_B + h X_C$.

The mutual information is calculated by calculating the probability $p\left(x | \alpha_k\right)$ of measuring $x \in \mathbb{C}$ at the output when coherent state $\alpha_k$ was sent, noting again that the realistic non-identical amplitudes and probabilities of the implemented QPSK alphabet may be readily included. Further details are found in Appendix~\ref{appendix:QSS}. 

Maximum QSS key rates are calculated from measured experimental parameters, Tab.~\ref{tab:lengths}. We see that twice the key rate, $2 \kappa$, is greater than the comparable key rate $\kappa$ for QKD-$f$ (remembering that one channel use is defined differently between QKD and QSS). In other words, QSS-$b$ outperforms pairwise QKD, by consuming fewer quantum resources. Protocol QSS-$b$ is therefore preferable over a classical information-theoretically secure protocol which can be performed over pairwise QKD-encrypted channels.

\begin{figure}[htp]
\centering
\includegraphics[width=\linewidth]{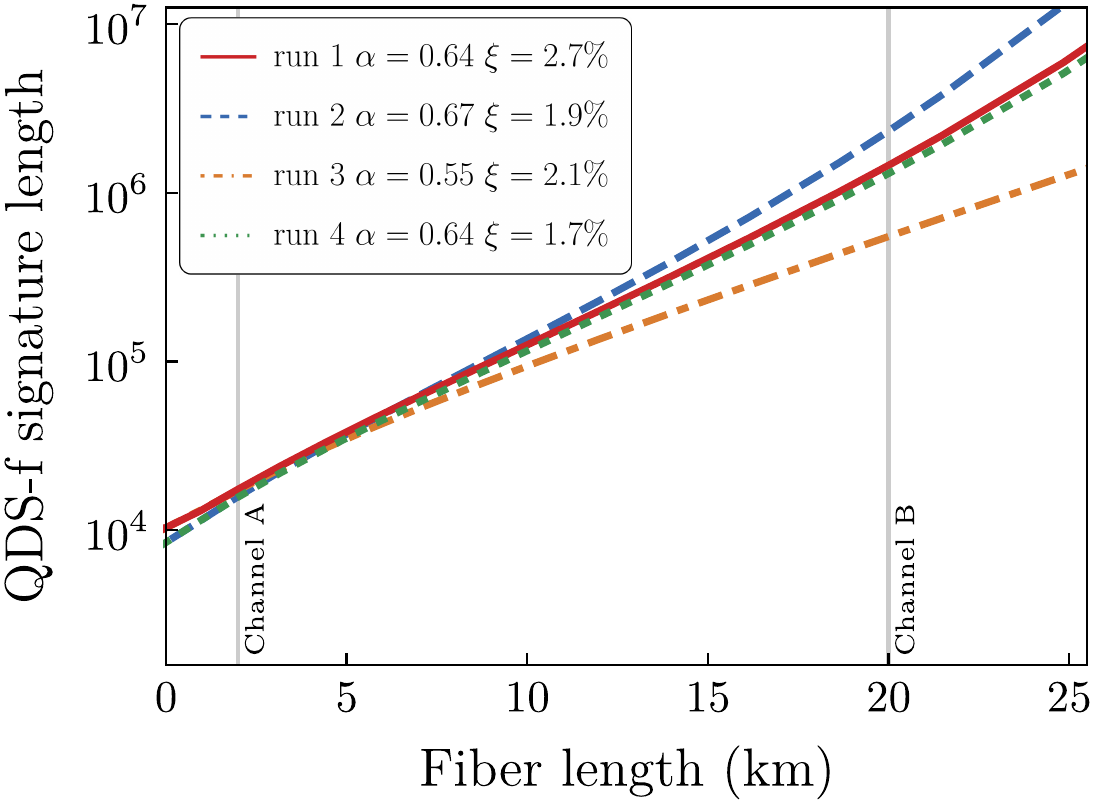}
\includegraphics[width=\linewidth]{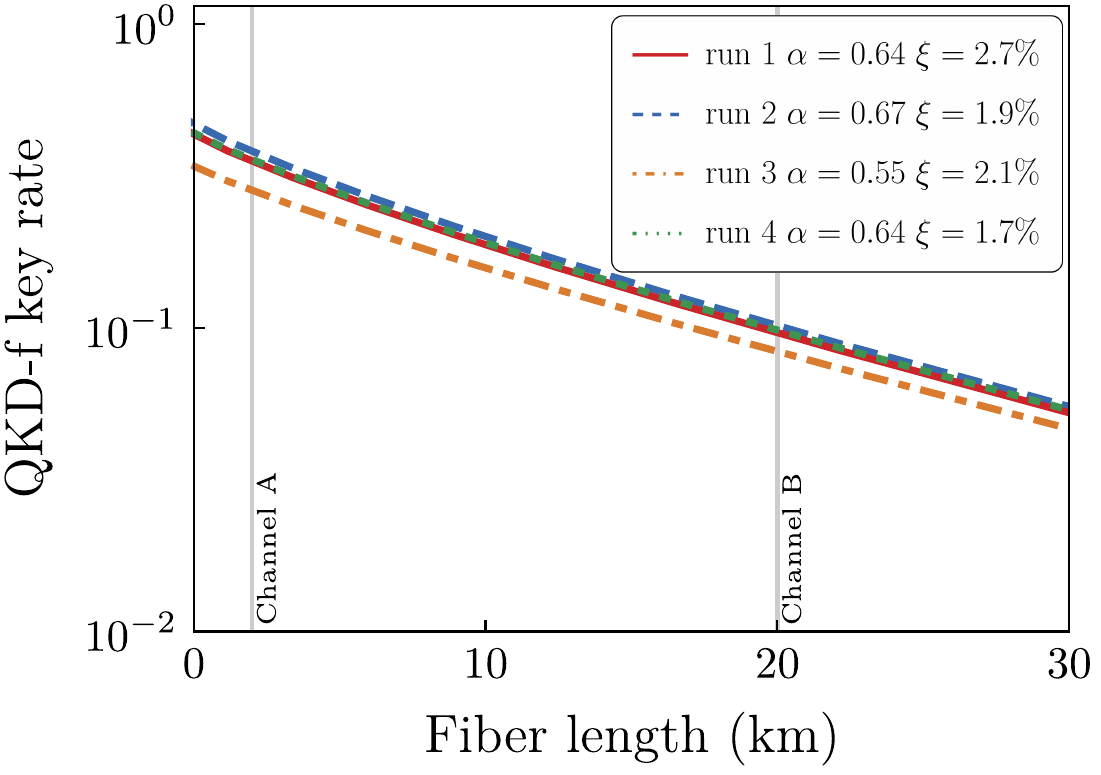}
\caption{\label{fig:qdsf_results} Calculated performance of agile and versatile system {\systemF}. Top: QDS signature lengths under protocol QDS-$f$ under a beamsplitter attack. Signature lengths $L$ at $20$~km (channel B) remain feasible with both ideal (above) and experimental realizations (Tab.~\ref{tab:lengths}). At $2$~km (channel A) the protocol requires small signature lengths and thus is the fastest QDS protocol over comparable distances, Fig.~\ref{fig:star}.
Bottom: Calculated maximum QKD key rates under protocol QKD-$f$ with a beamsplitter attack. Both: Vertical grid lines denote channel losses at which we have performed an experiment. Solid (red), dashed (blue), dot-dashed (orange) and dotted (green) lines correspond to experiments $1$, $2$, $3$ and $4$, respectively, and vertical grid lines depict loss levels over experimental channels $A$ and $B$, corresponding to fiber lengths $2$~km ($0.65$~dB loss) and $20$~km ($4.75$~dB loss). }
\end{figure}

\subsection{Second agile and versatile system {\systemF}}
For the second agile system, {\systemF}, Tx plays the role of Alice while Rx plays either Bob or Charlie.

\paragraph{QDS-$f$:} The performance of protocol QDS-$f$ is displayed in Fig.~\ref{fig:qdsf_results} under a beamsplitter attack. The excess noise and detector efficiency from experiment are included, and $p_e$ and $p_{err}$ are calculated via analogous methods to QDS-$b$, above. We see that in the ideal analysis of Fig.~\ref{fig:qdsf_results}, protocol QDS-$f$ allows for very small signature lengths $O\left(10^4\right)$ at $2$~km, while at $20$~km the predicted lengths are still very modest at $O\left(10^6\right)$.  

For small channel loss the required $L$ is roughly invariant over a broad range of $\alpha$, which suggests that QDS-$f$ is robust to experimental differences, and is thus easy to implement on an agile and versatile system alongside future alternative cryptographic protocols which may require more restrictive choice of $\alpha$. For large channel loss however the choice of $\alpha$ becomes increasingly important, but using for example the mean $\bar{\alpha}=0.55$ and $\xi=2.1\%$ from experimental run $3$, QDS-$f$ is predicted to remain secure even down to $20$~dB loss with still-feasible signature lengths $O\left(10^9\right)$, which would allow a one-bit message to be signed in approximately one second.

\begin{figure}[t!]
\centering
\includegraphics[width=\linewidth]{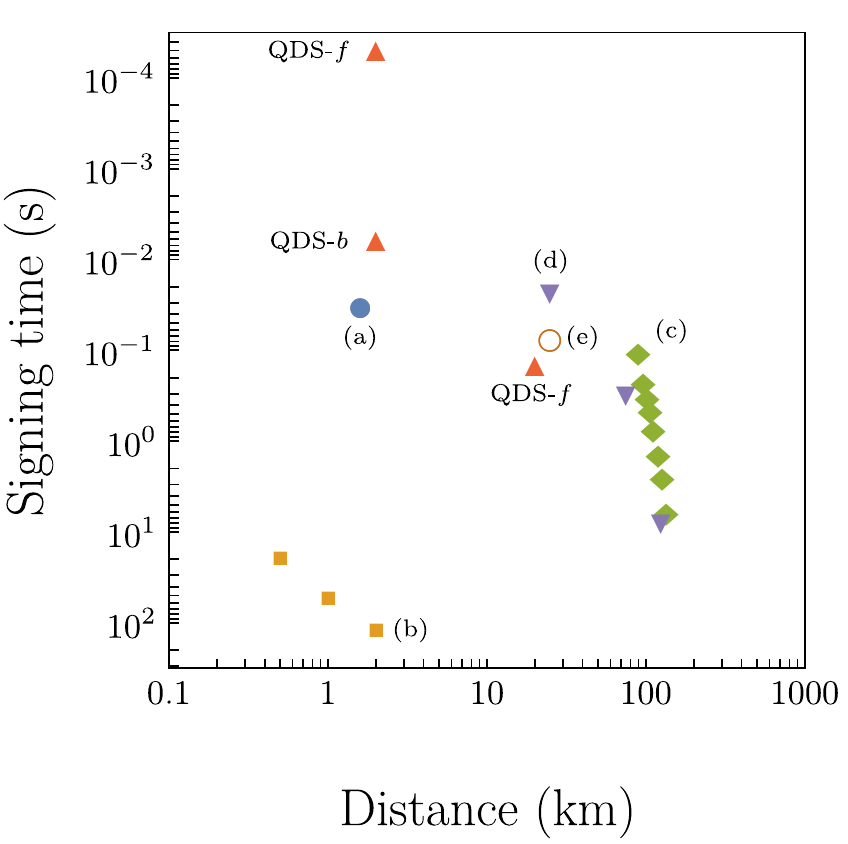}
\caption{\label{fig:star} Time required to sign a one-bit message, and the corresponding channel lengths, for several recent QDS protocols. At the short distances ($\sim 2$~km) favoured by the continuous-variable platform, our QDS-$f$ and QDS-$b$ allow for signing times of less than $0.05$~ms and $6$~ms, respectively, improving previous results in CV (a) and discrete-variable systems (b). At at $20$~km QDS-$f$ has signing time comparable to recent DV QDS systems (c)-(e). Protocols depicted: Red triangles - current paper. (a) Free-space CV QDS \cite{croal_free-space_2016}. (b) Unambiguous-state-elimination-based QDS \cite{donaldson_experimental_2016}. (c) Differential-phase-shift-based QDS \cite{collins_experimental_2016}. (d) GHz BB$84$ QDS \cite{an_practical_2019}. (e) Early QDS-QKD ``agile'' system with measurement-device-independent capabilities \cite{roberts_experimental_2017}.}
\end{figure}

A more realistic signature length may be calculated by using the $p_{err}$ directly measured from the output of Rx, which includes all noise sources and detector inefficiencies. This results in the signature lengths which are displayed in Tab.~\ref{tab:lengths}. Crucially, they remain highly feasible over the metropolitan distances where continuous-variable cryptography is expected to be effective. Of particular note is the $L = 47,887$ required to securely sign a $1$ bit message over $2$~km fiber, which to our knowledge makes QDS-$f$ the fastest ever demonstration of a QDS protocol, requiring just $0.047$~ms to sign a message at our $1$~GHz sending rate, Fig.~\ref{fig:star}.

\paragraph{QKD-$f$:} The calculated maximum secure key rates under protocol QKD-$f$ are plotted in Fig.~\ref{fig:qdsf_results} under a beamplitter attack. The performance of the protocol agrees with Ref.~\cite{Papanastasiou2018} and corroborates their results over comparable parameter regimes in amplitude, transmission and noise. The QPSK amplitudes reported in our experiment, however, are close to optimal. Calculated maximum key rates, deduced from experimental parameters, are displayed in Tab.~\ref{tab:lengths}. 

Finally, we want to note that key rates in a concrete implementation will depend on a number of parameters. For example, error correction in CV QKD can be computationally very demanding and will limit the obtainable key rate. An agile and versatile system therefore allows itself to resort to the protocol with the least demand on resources for a given task.

\section{Conclusion}\label{sec:6}
Agile and versatile quantum cryptography allows the introduction of a layer abstraction between the quantum optical hardware and the protocol layer based on firmware and software. This allows future quantum cryptography systems to be optimized towards agility and versatility and to explore how this concept can be applied to already existing ones. To underpin this concept, we have experimentally demonstrated versatility by showing that the same quantum sender and receiver can be utilized independent of the protocol run on top of it. Because the agile middleware is opaque to the user layer, by design, the inner workings of the task are inaccessible to the user. This implies agility.  The proposed layer abstraction could potentially be further developed through standardization groups \cite{etsi_isg_004, etsi_isg_014}. 

For the demonstrations we have utilized a continuous-variable quantum communication system that is almost exclusively built from commercial-off-the-shelf (COTS) telecom components. This makes it inherently compatible with telecom networks and allows C-band operation and high sending rates, since telecom components for coherent communication are optimized for GHz sending rates, even ranging up to 100 GHz as the state of the art. This setup has been operated at a sending rate of 1 GHz, however there is no known fundamental limit to these rates. The current limitation is the electronic noise of the coherent detection unit, which can be further optimized in future works.

The continuous-variable protocols investigated were quantum digital signatures (QDS), quantum secret sharing (QSS) and quantum key distribution (QKD). We have shown for the first time that postselection can be utilized for QDS, and have proven its enhanced robustness to noise and to channel loss. Postselection on QDS measurement outcomes decreases the required signature lengths and thus allows us to demonstrate the shortest signing time for realistic distances of $2$~km and signing times comparable to recent discrete-variable QDS protocols over $20$~km. Furthermore the security of the QDS protocols has been proven for forward and backward sending configurations, enabling them to be used in both of the agile and versatile systems presented in this paper.

\section*{Acknowledgments}\label{sec:7}

This research has received funding from the European Union's Horizon 2020 research and innovation programme under grant agreement No 820466 (CiViQ).
This research has also been funded by the German Federal Ministry of Education and Research (BMBF) within the project \emph{Hardware-based Quantum Security} (HQS).
The authors gratefully acknowledge the support from the Scottish Universities Physics Alliance (SUPA) and the Engineering and Physical Sciences Research Council (EPSRC). The project was supported within the framework of the International Max Planck Partnership (IMPP) with Scottish Universities.

\bibliography{AQC}

\appendix 

\section{QDS-$b$ security proof}\label{appendix:QDS2forge}
Recall that a QDS protocol must be secure against repudiation and forgery, and it should be robust and succeed if all players are honest. We will prove the security of our protocol against each of these, and finally derive Eq.~(\ref{eqn:QDSbL}), which implicitly defines the main figure of merit of a QDS protocol, the signature length $L$ required to sign a $1$~bit message. 

During a repudiation attack, Alice will try to cause Bob and Charlie to disagree about whether her message is genuine. Security against repudiation follows along similar lines to \cite{thornton_continuous-variable_2019, croal_free-space_2016, donaldson_experimental_2016}, and we reproduce key details below for completeness. 

We assume that Alice is free to manipulate her declared $A_{B, C}^m$, and she has full control over the mismatch rates $p_B \left(p_C\right)$ with respect to states which she originally sent to Bob or Charlie, and the $p_{B, C}$ may even be chosen to be zero.

After swapping, step three of the protocol, Bob and Charlie both possess two half-signatures, each of length $L/2$, consisting either of states which they held originally or which they received during swapping. Alice succeeds in her repudiation attack if Bob accepts both of his halves as genuine, and Charlie rejects at least one of his halves as fake. Therefore the probability of successful repudiation is given by

\begin{equation*}
\varepsilon_{rep} = P\left[\left(A \cap B \right) \cap \left( C \cup D \right) \right]
\end{equation*}

\noindent where $A \left(B\right)$ denotes the event that Bob accepts on his first (second) half, and $C \left(D\right)$ denotes the event that Charlie rejects on his first (second) half. Now, using probability inequalities $P\left(X \cap Y\right) \le min\left\{P\left(X\right), P\left(Y\right)\right\}$ and $P\left(X \cup Y\right) \le P\left(X\right) + P\left(Y\right)$ and Hoeffding's inequalities \cite{Hoeffding1963}, we see that $min\left\{P\left(A\right), P\left(B\right)\right\} \le \exp\left[ - \left(p - s_B\right)^2 L\right]$ and $P\left(C\right) + P\left(D\right) \le 2 \exp\left[-\left(s_C-p\right)^2L\right]$, where $p := max\left\{p_B, p_C\right\}$. 

Therefore we arrive at

\begin{align*}
\varepsilon_{rep} &\le min\left\{ 2 \exp\left[ - \left( p - s_B \right)^2 L\right], 2 \exp \left[ - \left( s_C - p\right)^2 L\right] \right\} \notag \\
& \le 2 \exp\left[ \frac{-\left(s_C - s_B\right)^2}{4} L\right]
\end{align*}

\noindent provided that $s_B < s_C$, and where in the second inequality we have taken $p = \left(s_B + s_C\right)/2$ in order to maximize $\varepsilon_{rep}$.

A QDS protocol is robust if it succeeds when all parties are honest. Even in this case there is a probability $p_{err}$ of mismatch, owing to the non-orthogonality of the QPSK alphabet. Since $s_B < s_C$, an honest message is more likely to be rejected by Bob than Charlie, so we bound this probability. The message will be rejected if Bob detects more than $s_B L/2$ mismatches on either half of his eliminated signature. Using Hoeffding's inequalities this occurs with probability

\begin{equation*}
\epsilon_{reject} \le 2 \exp\left[ - \left(s_B - p_{err}\right)^2 L\right]
\end{equation*}

\noindent provided that $s_B > p_{err}$, i.e. Bob's mismatch threshold is greater than the honest mismatch rate.

In a forging attack, an eavesdropper will aim to minimize their mismatch probability with respect to either of the $X_{B, C}^m$ generated by Bob and Charlie. Since Bob already knows half of $X_C^m$ (the information which Bob himself forwarded), and since $s_B < s_C$, the most dangerous forger is a dishonest Bob. He is therefore assumed to eavesdrop on Charlie's distribution of quantum states, and tries to gain information about the $L/2$ signature elements which Charlie generated himself.

Using Hoeffding's inequalities as in Ref.~\cite{thornton_continuous-variable_2019} we see that a forging attack succeeds with probability

\begin{equation*}
\varepsilon_{forg}  \le 2 \exp\left[ - \left(p_e - s_C\right)^2 \frac{L}{2}\right]
\end{equation*}

\noindent when $p_e > s_C$, and therefore all that is required is to bound $p_e$, which we now do.

Consider the $j^\text{th}$ signature element. Charlie holds some $c_j$ denoting which state from the QPSK alphabet he sent. Bob will declare an eliminated signature element, $B_j = \left\{b_j^1, b_j^2\right\}$ which is chosen to minimize $p_e$. The $b_j^1, b_j^2$ correspond to adjacent elements of the QPSK alphabet. A mismatch occurs if $b_j^1 = c_j$ or $b_j^2 = c_j$. Additionally, we assume that $B_j$ is the result of some optimal strategy involving Bob's quantum system, $\mathbb{B}_j$.

We define an error variable $E_j$ which takes the value $1$ if a mismatch occurs, and $0$ otherwise. Then $p_e \equiv P\left(E_j = 1\right)$, and the Shannon entropy $H\left(E_j\right) = h\left(p_e\right)$ is the binary entropy, since $|E_j| = 2$. Now, consider the conditional entropy $H\left(E_j, b_j^1, b_j^2 | c_j\right)$. Via the chain rule for conditional entropies,

\begin{equation*}
H\left(E_j, b_j^1, b_j^2 | c_j\right) = H\left(b_j^1, b_j^2 | c_j\right)
\end{equation*}

\noindent where we have used the fact that once $b_j^1, b_j^2$ and $c_j$ are known, $E_j$ is uniquely determined. Using the chain rule on $H\left(E_j, b_j^1, b_j^2 | c_j\right)$ again, but for a different variable, we get

\begin{align*}
H\left(E_j, b_j^1, b_j^2 | c_j\right) &= H\left(b_j^1, b_j^2 | E_j, c_j\right) + H\left(E_j | c_j\right) \notag \\
&\le H\left(b_j^1, b_j^2 | E_j, c_j\right) + h\left(p_e\right)
\end{align*}

\noindent since conditioning can never increase entropy. Therefore, by expanding the variable $E_j$,

\begin{align*}
H\left(b_j^1, b_j^2 | c_j\right) \le &\left(1 - p_e\right) H\left(b_j^1, b_j^2 | E_j = 0, c_j\right) \\ + &p_e H \left(b_j^1, b_j^2 | E_j=1, c_j\right) + h\left(p_e\right).
\end{align*}

\noindent Now, $H\left(b_j^1, b_j^2 | E_j=0, c_j\right) \le \log_2\left(2\right) = 1$, and similarly for $E_j=1$, and so

\begin{equation}\label{eqn:A7}
H\left(b_j^1, b_j^2 | c_j\right) \le 1 + h\left(p_e\right).
\end{equation}

\noindent Finally, we expand the conditional entropy in terms of the joint entropy and the mutual information,

\begin{align}\label{eqn:A8}
H\left(b_j^1, b_j^2 | c_j\right) &= H\left(b_j^1, b_j^2\right) - I\left(b_j^1, b_j^2 : c_j\right) \notag \\
&\ge 2 - \chi\left(b_j^1, b_j^2 : c_j\right)
\end{align}

\noindent where we have used the fact that \emph{a priori} there are four choices for the pair $b_j^1, b_j^2$, and where $\chi$ is the Holevo information \cite{Nielsen2010}. Combining Eqs.~(\ref{eqn:A7}),~(\ref{eqn:A8}) we arrive at Eq.~(\ref{eqn:qdsbsecurity}) from the main paper.

Once $p_e$ and $p_{err}$ are bounded for the protocol, the probability $\varepsilon_{fail}$ that the protocol fails can be found. For concreteness, we assign equal probability to the failure of the protocol either by allowing a forging or repudiation attack, or by aborting when all players are honest, that is

\begin{equation*}
\varepsilon_{fail} = \varepsilon_{forg} = \varepsilon_{rep} = \varepsilon_{reject}
\end{equation*}

\noindent and by choosing $s_B = p_{err} + \left(p_e + p_{err}\right)/4$; $s_C = p_{err} = 3\left(p_e - p_{err}\right)/4$, in order to satisfy the second two equalities, we arrive at Eq.~(\ref{eqn:QDSbL}) from the main paper

\begin{equation}
\varepsilon_{fail} \le 2 \exp \left[ - \frac{\left( p_e - p_{err} \right)^2}{16} L \right]
\end{equation}

\noindent when $p_{err} < s_B < s_C < p_e$.

Finally, we note that under a beamsplitter attack, Eve's \emph{a priori} state is
\begin{equation}
\rho_E = \sum_{k=0}^3 |\sqrt{1-T}\alpha_k\rangle\langle\sqrt{1-T}\alpha_k|
\end{equation}
when states $|\alpha_k\rangle$ from the QPSK alphabet are sent through lossy channel with transmittivity T. Eve's \emph{a posteriori} state is simply $\rho_{E}^k = |\sqrt{1-T}\alpha_k\rangle\langle \sqrt{1-T}\alpha_k|$, from which her Holevo information is calculated as

\begin{equation}
\chi = S\left(\rho_E\right) - \sum_{k=0}^3 p\left(k\right) S\left(\rho_E^k\right).
\end{equation}
with $S$ the Von Neuman entropy.

It is fitting to close this section with a brief discussion of the state of security proofs within the field of CV QDS. In security proofs for QDS we see the usual contrast between continuous-variable (CV) and discrete-variable (DV) platforms that is well-documented in the QKD literature. CV protocols offer a wide range of advantages in terms of implementations (speed, compatibility with infrastructure, etc.) but the security proofs are more involved than those on the DV platform, owing to the formidibly large Hilbert spaces. An advanced and general DV QDS securtiy analysis has been presented in Ref.~\cite{amiri_secure_2016}, which adapts a decoy-state BB$84$ QKD protocol to the task of QDS. Specifically, their paper leverages tight bounds for the smooth min-entropy to their QDS scecurity proof in order to provide security against coherent attacks. This has the additional advantage of providing security in the finite-size setting.

While security against coherent attacks has been proven for the decoy-state BB$84$ setup, other QDS platforms typically reach security levels which are comparible with the neighbouring QKD protocol. This is the case, for example, for the differential phase-shift QDS of Refs.~\cite{collins_experimental_2016, collins_experimental_2017}, and is the case for the fully continuous-variable QDS protocol presented here. QKD on the fully CV platform, relying on phase measurement of coherent states, has only recently been proven secure against general coherent attacks in the finite-size regime \cite{Leverrier2017}, and even then only for a Gaussian modulation of the coherent states. This choice of modulation is theoretically necessary---to simplify and bound the attack of the eavesdropper---but it is experimentally unrealistic. To our knowledge there is no full security proof for QPSK QKD which offers security against the general coherent attacks in the finite-size regime. Binary \cite{Zhao2009} and ternary \cite{Bradler2018} modulations have been asymptotically secured against collective attacks, but the bounds are not tight and the techniques are not expected to be generalizable to larger alphabets.

Recent QKD works \cite{Ghorai2019, Lin2019} have made advances towards full security with the QPSK alphabet by providing security against coherent attacks in the asymptotic limit. The first, Ref.~\cite{Ghorai2019}, provides security by relying on a small-amplitude assumption to ensure that the QPSK alphabet is close to a Gaussian modulation. They also make the assumption of Gaussian optimality; neither of these techniques are applicable to CV QDS. The second work, Ref.~\cite{Lin2019}, removes these assumptions and applies convex optimization methods to provide security, by building upon cutting-edge reformulations of the Devetak-Winter key rate bound and related semi-definite programming optimizations. Similar work should in the future be a key focus of the CV QDS community. We simply note here that should there become available a tight lower-bound for the eavesdropper's smooth min-entropy under the QPSK alphabet, we can readily insert it into our present QDS analysis with only minor modification to our proof.

\section{Postselection in CV QDS}\label{appendix:PS}

In the QKD context it has been known for some time that postselection will improve key rates in the presence of excess noise, and is even a requirement for distilling a key for $T < 1/2$ in the direct-reconciliation regime \cite{Silberhorn2002}. We are thus motivated to apply postselection to our QDS protocols in order to allow a message to be securely signed over a larger range of channel parameters.

To apply the postselection technique, recipients in the protocol will disregard unfavourable measurement outcomes, i.e. those for which a dishonest player is deemed to have too much knowledge, or for which the probability of honest mismatch is too high. We thus define a region $\mathcal{R}$ of phase-space, and only allow honest players to accept measurement outcomes $x \notin \mathcal{R}$. The region $\mathcal{R}$ is then varied to increase the range of channel parameters for which the QDS protocols are secure, and to minimize signature length $L$.
 
To be concrete, in this work we take $\mathcal{R}$ parameterized by $\Delta_r, \Delta_\theta$ in polar coordinates in phase-space. This is the same postselection region considered in the recent QKD work Ref.~\cite{Lin2019}, but if desired, more general regions may be readily considered. A protocol using no postselection technique may be retained by setting $\Delta_r \rightarrow 0$ and $\Delta_\theta \rightarrow 0$.

We will now consider how this application of postselection affects security of QDS-$b$ and QDS-$f$.

\subsection{QDS-$b$}
The crucial quantity which controls the security of a QDS protocol is $g_{sec} := p_e - p_{err}$, which intuitively describes how much worse a dishonest player should fare than an honest player. The protocol is secure provided that $g_{sec} > 0$. 

In QDS-$b$, $p_e$ does not depend on Alice's heterodyne measurement, since a dishonest player will attack the sender (Tx) of the quantum states, and so $p_e$ is unaffected by postselection. We thus calculate the transformation of $p_{err}$.

Although in an actual run of the protocol the honest mismatch rate $p_{err}$ should be estimated from a publicly disclosed subset of $A_{B, C}^m$ and $X_{B, C}^m$, it is illustrative to consider how $p_{err}$ may be calculated theoretically. When Charlie sends state $|\alpha_k\rangle$ through a lossy channel, transmittivity T, then Alice receives outcome $x \in \mathbb{C}$ with probability

\begin{equation}
p\left(x | \alpha_k\right) = \frac{1}{\pi} \exp\left( - | x - \sqrt{\frac{T}{2}} \alpha_k|^2\right).
\end{equation}

\noindent Thus, the probability of eliminating the state $|\alpha\rangle$, when no postselection is used, is

\begin{align}
p_{err} &= \int_0^\infty r \; \mathrm{d}r \int_{\pi/2}^{3 \pi/2} \mathrm{d}\theta \; p\left(r e^{i \theta} | \alpha\right) \notag \\
&= \frac{1}{2} \text{erfc}\left(\sqrt{\frac{T}{2}} |\alpha|\right).
\end{align}

\noindent Postselecting on the region $\mathcal{R}$, the mismatch probability becomes

\begin{align}\label{eqn:perrPS}
p_{err}\left(\Delta_r, \Delta_\theta\right) = \frac{1}{\mathcal{N}} \int_{\Delta_r}^\infty r \; \mathrm{d}r &\left( \int_{3 \pi/2 - \Delta_\theta}^{\pi + \Delta_\theta} \mathrm{d}\theta \; f\left(r, \theta\right) \right. \notag \\
&+\left. \int_{\pi/2+\Delta_\theta}^{\pi - \Delta_\theta} \mathrm{d}\theta \; f\left(r, \theta\right)\right)
\end{align}

\noindent with $f\left(r, \theta\right) = p\left(r e^{i \theta} | \alpha\right)$ and $\mathcal{N}$ the probability that the outcome $x \in \mathbb{C}$ is accepted, i.e. it falls within $\rcomp$. Probability $\mathcal{N}$ is calculated analogously to Eq.~(\ref{eqn:perrPS}).

For QDS-$b$ the probability $p_e$ does not depend on Alice's heterodyne measurement, since a dishonest player will attack the sender (Tx) of the quantum states. The dishonest player's \emph{a posteriori state} depends not on a recipient's heterodyne outcome, but only on the chosen distributed alphabet state, Appendix~\ref{appendix:QDS2forge}. The postselection technique alters the probability distribution of heterodyne measurement outcomes which are used in the protocol; coherent state sending probabilities remain uniform and unaffected. Therefore, for QDS-$b$, the dishonest mismatch probability $p_e$ is unaffected by the choice of postselection region. This will no longer be the case for QDS-$f$, and we discuss this in detail in the next section. 

Our analysis of the effects of postselection follows identically when excess noise is included, simply substituting in the requisite formulas from \cite{thornton_continuous-variable_2019, Papanastasiou2018}. Finally, we note that when postselection is used the signature length $L$ calculated from Eq.~(\ref{eqn:QDSbL}) should be rescaled in order to remain a useful figure of merit. While the normally calculated $L$ counts how many signature elements are required to sign the message, many of the states which were sent during the protocol will be rejected. Including the rejected states in our accounting, the figure of merit is rescaled as $L \rightarrow \tilde{L} := L/\mathcal{N}$. These $L$ and $\tilde{L}$ may now be directly compared between protocols, and so in Sec.~\ref{sec:5} we make no distinction between $L$ and $\tilde{L}$.

\subsection{QDS-$f$}
Postselection affects probability $p_{err}$ in the same way as it does under protocol QDS-$b$. For our protocol QDS-$f$, a dishonest player's declaration depends on an honest player's heterodyne outcome: so the dishonest mismatch probability $p_e$ must now also vary with $\mathcal{R}$. We recall that the key security result for QDS-$f$, taking dishonest Bob as the forger, is \cite{thornton_continuous-variable_2019}

\begin{equation}
h\left(p_e\right) \ge 1 - \chi
\end{equation}

\noindent with $\chi$ the Holevo information between Bob's quantum system and Charlie's eliminated signature element. For the $j^\text{th}$ signature element in QDS-$f$ this takes the form

\begin{equation}\label{eqn:qdsfholevo}
\chi\left(x_1^j, x_2^j : \mathbb{B}_j\right) = S\left(\rho_B^j\right) - \sum_{x_1^j, x_2^j} p\left(x_1^j, x_2^j\right) S\left(\rho_B^{x_1^j, x_2^j}\right).
\end{equation}

\noindent The \emph{a posteriori state} $\rho_B^{x_1^j, x_2^j}$ is the quantum state held by Bob when Charlie's eliminated signature element is $x_1^j, x_2^j$, and $\rho_B^j$ is Bob's \emph{a priori} state which is mixed over all eliminated signature elements.

Under the beamsplitter or entangling-cloner attacks considered in this work, the conditional state $\rho_{B|c}^j$ held by Bob after Charlie measures $c \in \mathbb{C}$ may be readily calculated as in \cite{thornton_continuous-variable_2019}. Then, since Charlie's eliminated signature element is entirely determined by the quadrant in which $c$ lies, the state $\rho_B^{x_1^j, x_2^j}$ is caclulated by mixing $\rho_{B|c}^j$ over an entire quandrant of phase-space

\begin{align}\label{eqn:quadmix}
\rho_B^{x_1^j, x_2^j} &= \frac{1}{\mathcal{N}} \int \rho_{B|c}^j \; \mathrm{d}^2 c \notag \\
& = 4 \int_{0}^\infty r \; \mathrm{d}r \int_{0}^{\pi/2} \rho_{B | r e^{i \theta}} \; \mathrm{d} \theta
\end{align}

\noindent where $\mathcal{N}$ is the required normalization factor, and where in the second line we have explicitly shown the calculation for a particular eliminated signature element. 

Then we see that when postselection over region $\mathcal{R}\left(\Delta_r, \Delta_\theta\right)$ is used, Eq.~(\ref{eqn:quadmix}) should be modified

\begin{equation}\label{eqn:aposterioriPS}
\rho_B^{x_1^j, x_2^j} = \frac{1}{\mathcal{N}} \int_{\Delta_r}^\infty r \; \mathrm{d}r \int_{\Delta_\theta}^{\pi/2 - \Delta_\theta} \rho_{B | r e^{i \theta}} \; \mathrm{d}\theta
\end{equation}

\noindent with $\mathcal{N}$ the same normalization factor as in Eq.~(\ref{eqn:perrPS}). The \emph{a priori} state is likewise found by mixing Eq.~(\ref{eqn:aposterioriPS}) over all quadrants, and thus $p_e$ may be calculated. The figure of merit for QDS-$f$ under postselection is now $\tilde{L}$, as in the preceding section, though since this may be directly compared with $L$ in the absence of postselection we make no distinction in the main body of the paper. All results presented have the optimal choice of $\mathcal{R}\left(\Delta_r\right)$, noting that variations in $\Delta_\theta$ provide only small changes to signature lengths in both QDS-$b$ and QDS-$f$, and so in the main body of the paper we set $\Delta_\theta=0$ in order to focus on the much larger effects of the radial variations.	


\section{QSS-$b$ security calculations}\label{appendix:QSS}
We will first demonstrate the security calculation of the protocol in the presence of an external eavesdropper, with Bob and Charlie assumed honest. The starting point for our calculation is

\begin{equation}\label{eqn:DW}
\kappa_{Eve} \ge I\left(X_B, X_C : X_A\right) - \chi\left(X_A : \mathbb{E}\right)
\end{equation}

\noindent denoting the maximum calculated key rate between the shared variable $X_B, X_C$ of Bob/Charlie, and Alice's $X_A := F\left(A_B, A_C\right)$. Eve's quantum system is denoted $\mathbb{E}$. 

Let $b, c$ denote elements from the QPSK alphabet which are sent by Bob/Charlie. Then the mutual information may be calculated once the probability $p\left(X_A = a | X_B = b, X_C = c\right)$, for Alice to receive element $X_A=a$ conditioned on particular QPSK states, is known. In the ideal case, probability $p\left(X_B=b, X_C=c\right) = 1/16$ since each of the QPSK states are equally likely, and so $p\left(X_B=b, X_C=c | X_A=a\right)$ may be calculated using Bayes' formula. Then, $H\left(X_B, X_C | X_A\right)$ is calculated by integrating $p\left(X_A=a\right) H\left(X_B, X_C | X_A=a\right)$ over all possible outcomes $a$, and finally
\begin{equation*}
I\left(X_B, X_C : X_A\right) = H\left(X_B, X_C\right) - H\left(X_B, X_C | X_A\right)
\end{equation*}

\noindent The Holevo term in Eq.~(\ref{eqn:DW}) may be calculated in the usual way from Eve's \emph{a priori} and \emph{a posteriori} states, with $\rho_E$ mixed over all $X_A$, and $\rho_E^a$ Eve's state when Alice holds $X_A=a$, with the channel modeled under either beamsplitter or entangling-cloner attacks.

Before the channel, the total Bob-Charlie state is $\rho_{B, C} = \rho_B \otimes \rho_C$, where in the ideal case each $\rho_B \left(\rho_C\right)$ are an equally weighted mixture over the QPSK alphabet. Passing through the channels, $\rho_{\mathbb{A}, \mathbb{E}} = \rho_{\mathbb{A}_B, \mathbb{E}_B} \otimes \rho_{\mathbb{A}_C, \mathbb{E}_C}$ where, for example, under a beamsplitter attack

\begin{align*}
\rho_{\mathbb{A}_B, \mathbb{E}_B} = \frac{1}{4} \sum_{k=0}^3 &|\sqrt{T}\beta_k\rangle_A\langle\sqrt{T}\beta_k| \notag \\
& \otimes | \sqrt{1-T} \beta_k\rangle_E\langle\sqrt{1-T}\beta_k|
\end{align*}
\noindent and similarly for $\rho_{\mathbb{A}_C, \mathbb{E}_C}$. Alice heterodynes on each of her modes and receives outcomes $A_B, A_C$.

Since the function $F$ is in general not injective, Eve's state is found by mixing $\rho_{\mathbb{E} | A_B} \otimes \rho_{\mathbb{E} | A_C}$ over Alice's measurement outcomes $A_B, A_C$ to reach the \emph{a posteriori} state $\rho_{\mathbb{E} | X_A}$. Finally, the \emph{a priori} state is given by

\begin{equation}
\rho_{\mathbb{E}} = \int \mathrm{d}^2X_A  \; P\left(X_A\right) \rho_{\mathbb{E} | X_A}
\end{equation}

\noindent and so Eve's Holevo information may be calculated. The Holevo information under an entangling-cloner attack is calculated analogously, but now the channel mixes the input $\rho_B, \rho_C$ with one arm of one of Eve's two entangled two-mode squeezed vacuum states. The remainder of the calculation proceeds identically, and is shown e.g. in Ref.~\cite{thornton_continuous-variable_2019} in the context of QDS.

Including a dishonest Bob, the key-rate reads

\begin{equation}\label{eqn:DW_dishonest-bob}
\kappa_{B} \ge I\left(X_A : X_B, X_C\right) - \chi\left(X_A : \mathbb{E} \mathbb{B}\right)
\end{equation}
where $\mathbb{E}\mathbb{B}$ is a quantum system shared between Bob and Eve. The main difference in calculation of both mutual information and Holevo information terms is that now the Eve-Bob conspiracy has knowledge about which state Bob sent, and so Bob's alphabet should no longer be mixed over.

A dishonest Charlie is taken into account identically, and the final key rate, including possibility for either Bob or Charlie to be dishonest, is given by \cite{Kogias2017}

\begin{equation}
\kappa \ge min \left\{\kappa_{B}, \kappa_{C}\right\}.
\end{equation}


\end{document}